\documentclass[10pt,journal,compsoc]{IEEEtran}

\usepackage{amsmath,amsfonts,amssymb}
\usepackage{mathtools}
\usepackage{amsthm}
\usepackage{subcaption}
\usepackage{alltt, xspace, times, epsfig, url}
\usepackage{gensymb,xfrac,array,booktabs,bm,xcolor}
\usepackage[linesnumbered,lined,ruled]{algorithm2e}
\usepackage{adjustbox}
\usepackage{booktabs}
\usepackage{soul}  

\DeclareMathOperator*{\argmax}{arg\,max}
\DeclareMathOperator{\EX}{\mathbb{E}}

\theoremstyle{definition}
\newtheorem{definition}{Definition}
\newtheorem{theorem}{Theorem}

\ifCLASSOPTIONcompsoc
  \usepackage[nocompress]{cite}
\else
  \usepackage{cite}
\fi

\begin{document}

\title{Secure and Utility-Aware Data Collection with Condensed Local Differential Privacy}

\author{Mehmet~Emre~Gursoy,~\IEEEmembership{Student~Member,~IEEE,}
        Acar~Tamersoy,
        Stacey~Truex,
        Wenqi~Wei,
        and Ling~Liu,~\IEEEmembership{Fellow,~IEEE}
\IEEEcompsocitemizethanks{\IEEEcompsocthanksitem M.E. Gursoy, S. Truex, W. Wei and L. Liu are with the College of Computing, Georgia Institute of Technology, Atlanta,
GA. E-mail: \{memregursoy, staceytruex, wenqiwei\}@gatech.edu, ling.liu@cc.gatech.edu
\IEEEcompsocthanksitem A. Tamersoy is with Symantec Research Labs, Culver City, CA.\protect\\ E-mail: acar\_tamersoy@symantec.com}
\thanks{Manuscript accepted for publication in IEEE TDSC.}}

\markboth{IEEE Transactions on Dependable and Secure Computing,~Vol.~X, No.~Y, 2019}%
{}

\IEEEtitleabstractindextext{%
\begin{abstract}
Local Differential Privacy (LDP) is popularly used in practice for privacy-preserving data collection. Although existing LDP protocols offer high utility for large user populations (100,000 or more users), they perform poorly in scenarios with small user populations (such as those in the cybersecurity domain) and lack perturbation mechanisms that are effective for both ordinal and non-ordinal item sequences while protecting sequence length and content simultaneously. In this paper, we address the small user population problem by introducing the concept of Condensed Local Differential Privacy (CLDP) as a specialization of LDP, and develop a suite of CLDP protocols that offer desirable statistical utility while preserving privacy. Our protocols support different types of client data, ranging from ordinal data types in finite metric spaces (numeric malware infection statistics), to non-ordinal items (OS versions, transaction categories), and to sequences of ordinal and non-ordinal items. Extensive experiments are conducted on multiple datasets, including datasets that are an order of magnitude smaller than those used in existing approaches, which show that proposed CLDP protocols yield high utility. Furthermore, case studies with Symantec datasets demonstrate that our protocols accurately support key cybersecurity-focused tasks of detecting ransomware outbreaks, identifying targeted and vulnerable OSs, and inspecting suspicious activities on infected machines. 
\end{abstract}

}

\maketitle

\IEEEdisplaynontitleabstractindextext
\IEEEpeerreviewmaketitle

\ifCLASSOPTIONcompsoc
\IEEEraisesectionheading{\section{Introduction} \label{sec:introduction}}
\else
\section{Introduction} \label{sec:introduction}
\fi

\newlength{\textfloatsepsave} \setlength{\textfloatsepsave}{\textfloatsep}

\IEEEPARstart{O}rganizations and companies are becoming increasingly interested in collecting user data and telemetry to make data-driven decisions. While collecting and analyzing user data is beneficial to improve services and products, users' privacy poses a major concern. Recently, the concept of \textit{Local Differential Privacy (LDP)} has emerged as the accepted standard for privacy-preserving data collection \cite{duchi2013local,erlingsson2014rappor,wang2017locally}. In LDP, each user locally perturbs their sensitive data on their device before sharing the perturbed version with the data collector. The perturbation is performed systematically such that the data collector cannot infer with strong confidence the true value of any user given their perturbed value, yet it can still make accurate inferences pertaining to the general population. Due to its desirable properties, LDP has been adopted by major companies to perform certain tasks, including Google to analyze browser homepages and default search engines in Chrome~\cite{erlingsson2014rappor,fanti2016building}, Apple for determining emoji frequencies and spelling prediction in iOS~\cite{thakurta2017learning,thakurta2017emoji}, and Microsoft to collect application telemetry in Windows 10~\cite{ding2017collecting}.

While LDP is popularly used for the aforementioned purposes, one domain that is yet to embrace it is cybersecurity. It can be argued that this nuanced domain has the potential to greatly benefit from an LDP-like protection mechanism. This is because many security products rely on information collected from their clients, with the required telemetry ranging from file occurrence information in file reputation systems~\cite{chau2011polonium,karampatziakis2012using,tamersoy2014guilt} to heterogeneous security event information such as system calls and memory dumps in the context of Endpoint Detection and Response systems (see~\cite{egele2012survey} for a survey on core behavioral detection techniques used by such systems). Nevertheless, clients are often reluctant to share such data fearing that it may reveal the applications they are running, the files they store, or the overall cyber hygiene of their devices. Offering formal privacy guarantees would help convince clients that sharing their data will not cause privacy leakages.

On the other hand, existing LDP protocols suffer from problems that hinder their deployment in the cybersecurity domain. First, although they are accurate for large populations (e.g., hundreds of thousands of clients), their accuracy suffers when client populations are smaller. Population size is not necessarily a problem for the likes of Google Chrome and Apple iOS with millions of active users, but it does cause problems in a domain such as cybersecurity where small population sizes are common. For example, if a security analyst is analyzing the behavior of a particular malware that targets a certain system or vulnerability, only those clients who are infected by the malware will have meaningful observations to report, but the number of infections could be limited to less than a couple of thousand users globally due to the targeted nature of the malware. In such cases, we need a new scheme that allows the security analyst to make accurate inferences while simultaneously giving adequate privacy to end users. Second, existing protocols consider a limited set of primitive data types. To the best of our knowledge, currently no protocol supports perturbation of item sequences (with either ordinal or non-ordinal item domains) to offer privacy with respect to sequence length and content simultaneously. Sequences are more difficult to handle compared to the data types of singleton items or itemsets since they are not only high-dimensional, but also contain an ordering that must be preserved for tasks such as pattern mining. Yet, sequential data is ubiquitous in cybersecurity, e.g., security logs, network traffic data, file downloads, and so forth are all examples of sequential data.

In this paper, we propose the notion of \textit{Condensed LDP (CLDP)} and a suite of protocols satisfying CLDP to tackle these issues. In practice, CLDP is similar to LDP with the addition of a \textit{condensation} aspect, i.e., during the process of perturbation, similar outputs are systematically favored compared to distant outputs using condensed probability. We design protocols satisfying CLDP for various types of data, including singletons and sequences of ordinal and non-ordinal items. We show that CLDP can be satisfied by a variant of the Exponential Mechanism \cite{mcsherry2007mechanism}, and employ this mechanism as a building block in our \textit{Ordinal-CLDP}, \textit{Item-CLDP}, and \textit{Sequence-CLDP} protocols. Our methods are generic and can be easily applied for privacy-preserving data collection in multiple domains, including but not limited to cybersecurity. 

We consider a Bayesian adversary model for privacy-preser\-ving data collection, which enables us to establish a formal connection between LDP and CLDP and ensures that they provide the same protection against this common adversary. The Bayesian model measures the adversary's maximum posterior confidence (MPC) in predicting the user's true value over all possible inputs and outputs of a perturbation mechanism, where higher adversarial confidence implies lower privacy protection for users. Under this Bayesian model, we derive how the parameters of CLDP protocols can be selected to give as strong protection as LDP protocols. Experiments conducted using this parameter selection show that our CLDP protocols provide high utility by yielding accurate insights for population sizes that are an order of magnitude smaller than those currently assumed by state-of-the-art LDP approaches.

We also perform extensive experiments to evaluate the effectiveness of our protocols on real-world case studies and public datasets. Experiments show that proposed CLDP protocols enable accurate frequency estimation, heavy hitter identification, and pattern mining while achieving strong protection against the adversary model considered. Using data from Symantec, a major cybersecurity vendor, we show that CLDP can be used in practice for use cases involving ransomware outbreak detection, OS vulnerability analysis, and inspecting suspicious activities on infected machines. In contrast, existing LDP protocols can either not be applied to these problems or their application yields unacceptable accuracy loss. To the best of our knowledge, our work is the first to apply the concept of local differential privacy to the nuanced domain of cybersecurity. 

\vspace{-5pt}
\section{Background and Problem Setting}

We consider the data collection setting, where there are many clients (\textit{users}) and an untrusted data collector (\textit{server}). Each client possesses a secret value. The client's secret value can be an ordinal item (e.g., numeric value or integer), a categorical item, a non-ordinal item, or a sequence of items. The server wants to collect data from clients to derive useful insights; however, since the clients do not trust the server, they perturb their secrets locally on their device before sharing the perturbed version with the server. Randomized perturbation ensures that the server, having observed the perturbed data, cannot infer the true value of any one client with strong probability. At the same time, the scheme allows the server to derive useful insights from \textit{aggregates} of perturbed data by recovering statistics pertaining to the general population.

In Section \ref{sec:threatmodel}, we start by introducing the threat and Bayesian adversary models we consider for measuring privacy protection in the above data collection setting. In Section \ref{sec:ldp}, we define LDP as the current solution to privacy-preserving data collection. In Section \ref{sec:utilitymodel}, we describe the utility model and provide preliminary analysis that shows LDP's utility loss under small user populations. Section \ref{sec:problemstatement} contains our problem statement.

\vspace{-5pt}
\subsection{Adversary Model} \label{sec:threatmodel}

We use a Bayesian approach to measure privacy in the local data collection setting. Our Bayesian adversary model is similar to the Bayesian adversary formulations for membership privacy and location privacy in \cite{li2013membership,yang2015bayesian,kasiviswanathan2014semantics,yu2017dynamic}; but differs in the sense that the main goal of local privacy schemes is to offer confidentiality and plausible deniability for each individual user's secret. Higher plausible deniability implies lower adversarial prediction confidence and consequently higher privacy protection.

The threat stems from an untrusted third party (e.g., the data collector) inferring the true value of the user with high confidence from the perturbed value (s)he observes. Then, the goal of randomized perturbation is to stop an adversary $\mathcal{A}$, even if they can fully observe perturbed output $y$, from inferring the user's true secret $v$. Given $y$, the optimal attack strategy for $\mathcal{A}$ is:
\begin{align}
\mathcal{A}(y) &= \argmax_{\hat{v} \in \mathcal{U}} \text{Pr}[\hat{v}|y] \\ 
&= \argmax_{\hat{v} \in \mathcal{U}}  \frac{\pi(\hat{v}) \cdot \text{Pr}[f(\hat{v}) = y]} {\sum_{z \in \mathcal{U}} \pi(z) \cdot \text{Pr}[f(z) = y]}
\end{align}
where $\pi(\hat{v})$ denotes the prior probability of $\hat{v}$, $f$ denotes a perturbation function, and $\mathcal{U}$ denotes the universe of possible items. Then, worst-case privacy disclosure can be measured using the maximum posterior confidence (MPC) the adversary can achieve over all possible inputs and outputs:
\begin{align} \label{eq:mpc}
\text{MPC} = \max_{v \in \mathcal{U}, y} \text{Pr}[v|y] = \max_{v \in \mathcal{U}, y} \frac{\pi(v) \cdot \text{Pr}[f(v) = y]} {\sum_{z \in \mathcal{U}} \pi(z) \cdot \text{Pr}[f(z) = y]}
\end{align}
This establishes a mathematical framework under which we can quantify worst-case adversarial disclosure of perturbation protocols, for both informed adversaries (with prior knowledge $\pi$) and uninformed adversaries (e.g., by canceling out the $\pi$ term, or equivalently, setting $\pi(x) = 1/|\mathcal{U}|$ for all $x \in \mathcal{U}$). When MPC is high, the adversary can more easily predict the true input of the user, hence $f$ yields lower privacy. Thus, lower MPC is desired.

\vspace{-5pt}
\subsection{Local Differential Privacy} \label{sec:ldp}

The state-of-the-art scheme currently used and deployed by major companies such as Google, Apple, and Microsoft in the data collection setting is LDP \cite{erlingsson2014rappor,fanti2016building,ding2017collecting,thakurta2017emoji,thakurta2017learning}. In LDP, each user perturbs their true value $v$ using an algorithm $\Psi$ and sends $\Psi(v)$ to the server. LDP can be formalized as follows.

\begin{definition}[$\varepsilon$-LDP] \label{def:LDP}
A randomized algorithm $\Psi$ satisfies $\varepsilon$-local differential privacy ($\varepsilon$-LDP), where $\varepsilon > 0$, if and only if for any inputs $v_1,v_2$ in universe $\mathcal{U}$, we have:
\[
\forall y \in Range(\Psi): ~~ \frac{\text{Pr}[\Psi(v_1) = y]}{\text{Pr}[\Psi(v_2) = y]} \leq e^{\varepsilon}
\]
where $Range(\Psi)$ denotes the set of all possible outputs of algorithm $\Psi$. 
\end{definition}

Here, $\varepsilon$ is the privacy parameter controlling the level of indistinguishability. Lower $\varepsilon$ yields higher privacy. Several works were devoted to building accurate protocols that satisfy $\varepsilon$-LDP. These works were analyzed and compared in \cite{wang2017locally}, and it was found that the two currently optimal protocols are based on: (i) OLH, the hashing extension of the GRR primitive and (ii) RAPPOR, the Bloom filter-based bitvector encoding and bit flipping strategy. Next, we briefly present these protocols.

\vspace{2pt}
\noindent
\textbf{Generalized Randomized Response (GRR).}
This protocol is a generalization of the \textit{Randomized Response} survey technique introduced in \cite{warner1965randomized}. Given the user's true value $v$, the perturbation function $\Psi_{GRR}$ outputs $y$ with probability:
\[
\text{Pr}[\Psi_{GRR}(v)=y] = 
\begin{cases}
p=\frac{e^{\varepsilon}}{e^{\varepsilon}+|\mathcal{U}|-1} & \text{ if } y=v \\
q=\frac{1}{e^{\varepsilon}+|\mathcal{U}|-1} & \text{ if } y \neq v
\end{cases}
\]
where $|\mathcal{U}|$ denotes the size of the universe. This satisfies $\varepsilon$-LDP since $\frac{p}{q}=e^{\varepsilon}$. In words, $\Psi_{GRR}$ takes as input $v$ and assigns a higher probability $p$ to returning the same output $y=v$. With remaining $1-p$ probability, $\Psi_{GRR}$ samples a fake item from the universe $\mathcal{U} \setminus \{v\}$ uniformly at random, and outputs this fake item. 

\vspace{2pt}
\noindent
\textbf{Optimized Local Hashing (OLH). }
When the universe size $|\mathcal{U}|$ is large, it dominates the denominator of $p$ and $q$, thus the accuracy of GRR deteriorates quickly. The OLH protocol proposed in \cite{wang2017locally} handles the large universe problem by first using a hash function to map $v$ into a smaller domain of hash values and then applying GRR on the hashed value. Formally, the client reports: 
\[
\Psi_{OLH}(v) = \langle H, \Psi_{GRR}(H(v)) \rangle
\]
where $H$ is a randomly chosen hash function from a family of hash functions. Each hash function in the family maps $v \in \mathcal{U}$ to a domain $\{1...g\}$, where $g$ denotes the size of the hashed domain, typically $g \ll |\mathcal{U}|$. We use $g = \lceil e^{\varepsilon} + 1 \rceil$ as the optimal value of $g$ found in \cite{wang2017locally}.

\vspace{2pt}
\noindent
\textbf{Randomized Aggregatable Privacy-Preserving Ordinal Response (RAPPOR). }
RAPPOR was developed by Google and is used in Chrome \cite{erlingsson2014rappor}. In RAPPOR, the user's true value $v$ is encoded in a bitvector $B$. The straightforward method is to use one-hot encoding such that $B$ is a length-$|\mathcal{U}|$ binary vector where the $v$'th position is 1 and the remaining positions are 0. When $|\mathcal{U}|$ is large, both communication cost and inaccuracy cause problems, hence RAPPOR uses Bloom filter encoding. Specifically, $B$ is treated as a Bloom filter and a set of hash functions $\mathcal{H}$ is used to map $v$ into a set of integer positions that must be set to 1. That is, $\forall H \in \mathcal{H}$, $B[H(v)]=1$, and the remaining positions are 0. 

After the encoding, RAPPOR uses perturbation function $\Psi_{RAPPOR}$ on $B$ to obtain perturbed bitvector $B'$ as follows:
\[
\text{Pr}\Big[B'[i]=\Psi_{RAPPOR}(B[i])=1\Big] = 
\begin{cases}
\frac{e^{\varepsilon/2\Delta}}{e^{\varepsilon/2\Delta}+1} & \text{ if } B[i]=1 \\
\frac{1}{e^{\varepsilon/2\Delta}+1} & \text{ if } B[i]=0
\end{cases}
\]
where $2\Delta$ is analogous to the notion of \textit{sensitivity} in differential privacy \cite{inan2017sensitivity}, i.e., how many positions can change in neighboring bitvectors at most? In one-hot encoding, $\Delta=1$; in Bloom filter encoding, $\Delta=|\mathcal{H}|$. Then, the perturbation process considers each position in $B$ independently, and the existing bit $B[i]$ is either kept or flipped when creating $B'[i]$. 

\vspace{-6pt}
\subsection{Utility Model and Analysis} \label{sec:utilitymodel}

The most common use of LDP is to enable the data collector learn \textit{aggregate} population statistics from large collections of perturbed data. Much research has been invested in tasks such as frequency estimation (identify proportion of users who have a certain item) and heavy hitter discovery (identify popular items that are held by largest number of users) \cite{wang2017locally,bassily2015local,bassily2017practical,qin2016heavy}. Utility is measured by how closely the privately collected statistics resemble the actual statistics that would have been obtained if privacy was not applied.

Frequency estimation and heavy hitter discovery are important tasks in the cybersecurity domain as well. For example, by monitoring the observed frequencies of different malware using aggregates of privatized malware reports, a cybersecurity vendor such as Symantec can identify large-scale malware outbreaks and create response teams to address them. Furthermore, analyzing the heavy hitter operating systems that are most commonly infected by the malware will enable Symantec understand OS vulnerabilities as well as the fraction of clients in its user base that are impacted. Relevant findings may also be used by Symantec when developing its next-generation anti-malware defenses. 

\begin{figure}[!t]
    \centering
    \includegraphics[width=.43\textwidth]{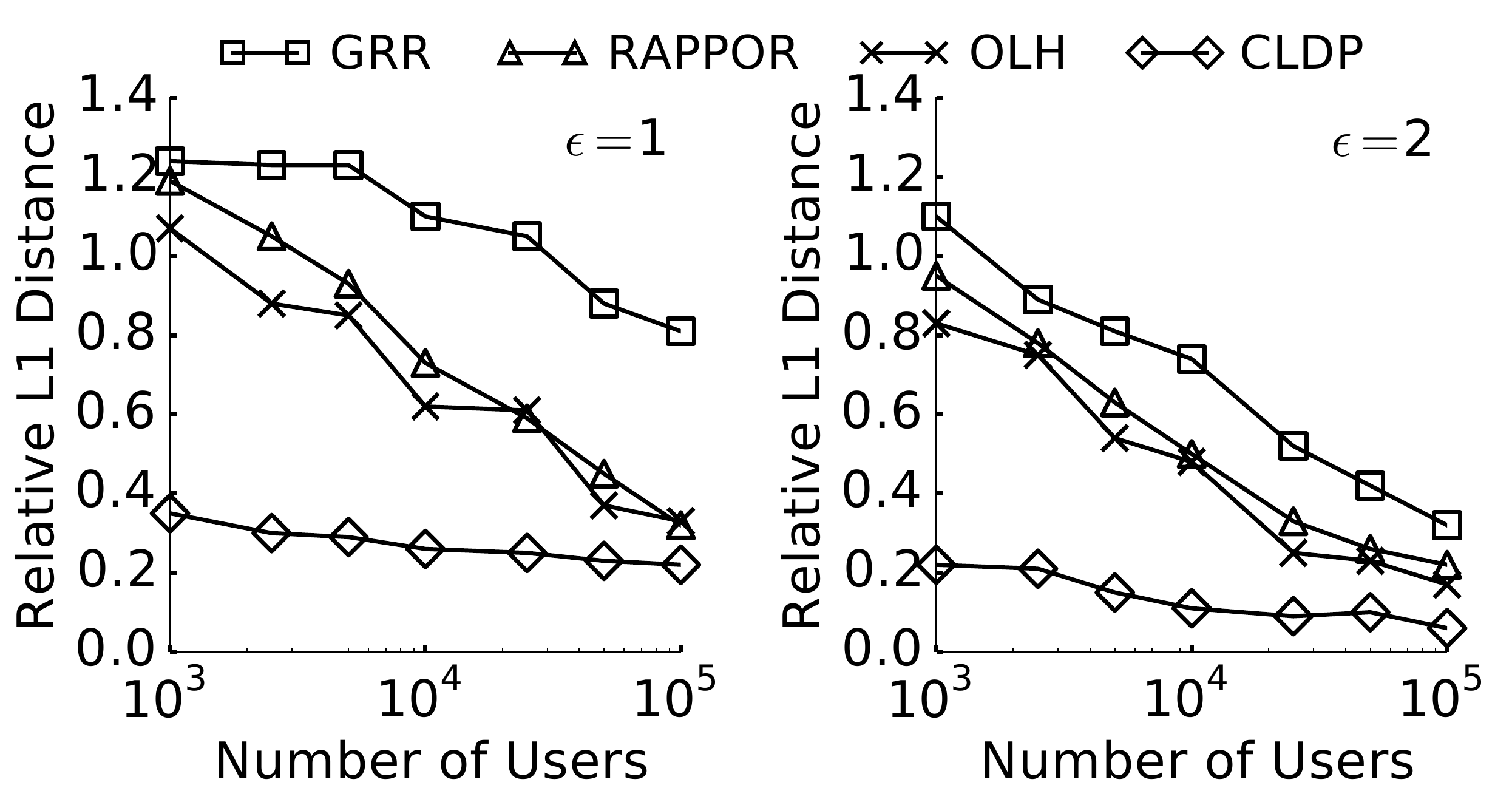}
    \vspace{-8pt}
    \caption{Measuring relative error in item frequency estimation by calculating L1 distance between actual and estimated frequencies. L1 distance of $d$ means $100 \cdot d$ \% estimation error. 
    }
    \label{fig:L1users}
    \vspace{-12pt}
\end{figure}

A novel challenge posed by the cybersecurity domain, however, is accurately supporting small user populations. Typically, existing LDP literature assumes the availability of ``large enough" user populations in the order of hundreds of thousands or millions of users \cite{erlingsson2014rappor,wang2017locally,qin2016heavy,thakurta2017learning}. Yet, population sizes in the cybersecurity domain are typically much smaller. For example, consider a security analyst analyzing the behavior of a specific malware by studying infected user machines. It is often the case that malware targets a specific computing platform or software product, limiting the total number of infections to less than a couple of thousand users globally. We designed the frequency estimation experiment in Figure~\ref{fig:L1users} to illustrate how the utility of existing LDP protocols suffer under such small populations. We sampled each user's secret value from a Gaussian distribution with mean $\mu$~=~50 and standard deviation $\sigma$~=~12, and rounded it to the nearest integer. The goal of the data collector is to estimate the true frequency of each integer. We run this experiment for varying number of users between 1,000 and 100,000 and graph the error in the frequency estimations made by the data collector.

We observe from Figure \ref{fig:L1users} that although more recent and optimized protocols improve previous ones by decreasing estimation error (e.g., OLH outperforms RAPPOR, which outperforms GRR); in cases with small user populations (e.g., 1000, 2500, or 5000 users) the improvements offered by more recent LDP protocols over previous ones are only 10-20\%, whereas our proposed CLDP approach provides a remarkable 60-70\% improvement. Furthermore, with 2500 users, estimation error is larger than 80\% even for the state-of-the-art OLH algorithm, which is optimized for frequency estimation \cite{wang2017locally}. In contrast, our proposed CLDP solution is able to handle small user populations gracefully, with estimation errors lower than half of OLH's errors.

\vspace{-6pt}

\subsection{Problem Statement} \label{sec:problemstatement}

As our analysis shows, low utility levels of existing LDP protocols under small user population sizes constitute an important obstacle towards their deployment in the cybersecurity domain. This motivates us to seek alternative approaches and protocols for privacy-preserving data collection in this challenging scenario. The problem we study in this paper is to design a data collection scheme such that: (i) it gives at least as strong privacy protection as existing LDP protocols under the MPC adversary model, (ii) while doing so, it provides higher accuracy and data utility compared to existing protocols especially for small user populations, and (iii) it offers extensibility and generalizability to support complex data types such as different types of singleton items, itemsets, and sequences that are present in the cybersecurity domain.

\vspace{-4pt}
\section{Proposed Solution}

In this section, we introduce our proposed solution aproach to the problem stated above. We start with the notion of \textit{Condensed} Local Differential Privacy (CLDP).

\vspace{-4pt}
\subsection{Condensed Local Differential Privacy} \label{sec:cldp}

Let $\mathcal{U}$ denote the finite universe of possible values (items) and let $d: \mathcal{U} \times \mathcal{U} \rightarrow [0,\infty)$ be a distance function that takes as input two items $v_1,v_2 \in \mathcal{U}$ and measures their distance. We require $d$ to satisfy the conditions for being a metric, i.e., non-negativity, symmetry, triangle inequality, and identity of discernibles. Then, CLDP can be formalized as follows.

\begin{definition}[$\alpha$-CLDP] \label{def:CLDP}
A randomized algorithm $\Phi$ satisfies $\alpha$-condensed local differential privacy ($\alpha$-CLDP), where $\alpha > 0$, if and only if for any inputs $v_1,v_2 \in \mathcal{U}$:
\[
\forall y \in Range(\Phi): ~~ \frac{\text{Pr}[\Phi(v_1) = y]}{\text{Pr}[\Phi(v_2) = y]} \leq e^{\alpha \cdot d(v_1,v_2)}
\]
where $Range(\Phi)$ denotes the set of all possible outputs of algorithm $\Phi$.
\end{definition}

LDP and CLDP follow a similar formalism, but differ in how their privacy parameters and indistinguishability properties work. Similar to $\varepsilon$-DP, $\alpha$-CLDP satisfies the property that an adversary observing $y$ will not be able to distinguish whether the original value was $v_1$ or $v_2$. However, in $\alpha$-CLDP, indistinguishability is controlled also by items' distance $d(\cdot,\cdot)$ in addition to $\alpha$. Consequently, as $d$ increases, $\alpha$ must decrease to compensate, i.e., we have $\alpha \ll \varepsilon$. By definition, CLDP constitutes a metric-based extension of LDP. Metric-based extensions of differential privacy have been studied in the past under certain settings such as aggregate query answering in centralized statistical databases \cite{chatzikokolakis2013broadening}, geo-indistinguishability in location-based systems \cite{andres2013geo,bordenabe2014optimal}, and protecting sensitive relationships between entities in graphs through $k$-edge differential privacy \cite{hay2009accurate}. In contrast, we propose the metric-based CLDP extension in the \textit{data collection} setting. Our data collection setting poses novel challenges due to: (i) the local privacy scenario, unlike centralized DP assumed in aggregate query answering and graph mining in which user data is collected in the clear first and privacy is applied after the data has been stored in a centralized database, (ii) data types that are different than tabular datasets, locations, and graphs, and (iii) establishing relationships and comparison between CLDP and LDP under the assumed adversary and utility models.

Since existing LDP protocols do not satisfy CLDP, we need new mechanisms and protocols supporting CLDP. We show below that a variant of the Exponential Mechanism (EM) \cite{mcsherry2007mechanism} satisfies $\alpha$-CLDP. EM is used in the remainder of the paper as a building block for more advanced CLDP protocols. 

\vspace{2pt}
\noindent
\textbf{Exponential Mechanism (EM). }
Let $v \in \mathcal{U}$ be the user's true value, and let the Exponential Mechanism, denoted by $\Phi_{EM}$, take as input $v$ and output a perturbed value in $\mathcal{U}$, i.e., $\Phi_{EM}: \mathcal{U} \rightarrow \mathcal{U}$. Then, $\Phi_{EM}$ that produces output $y$ with the following probability satisfies $\alpha$-CLDP:
\[
\forall y \in \mathcal{U}: ~~ \text{Pr}[\Phi_{EM}(v) = y] = \frac{e^{\frac{-\alpha \cdot d(v,y)}{2}}}{\sum_{z \in \mathcal{U}} e^{\frac{-\alpha \cdot d(v,z)}{2}}}
\]

\begin{theorem} \label{thm:EM}
Exponential Mechanism satisfies $\alpha$-CLDP.
\end{theorem}
\vspace{-10pt}
\begin{proof}
Provided in the appendix.
\end{proof}

\vspace{-10pt}
\subsection{Privacy Protection of LDP and CLDP} \label{sec:conversion}
\vspace{-1pt}

We wish to find the appropriate $\alpha$ value to be used in CLDP so that the MPC under $\alpha$-CLDP will be equal to or lower than the MPC under $\varepsilon$-LDP, which ensures that CLDP gives equal or better protection than LDP against the adversary we consider.

\vspace{2pt}
\noindent\textbf{Question:} Let $\mathcal{U}$, $d$ and $\pi$ be given. If there is an LDP protocol currently in place with privacy budget $\varepsilon$ and we are interested in switching to $\alpha$-CLDP, how should the value of $\alpha$ be selected to achieve equal or better protection than LDP according to the MPC adversary model? 

\vspace{2pt}
\noindent\textbf{Answer:} Based on the quantification of the adversary's maximum posterior confidence in Equation \ref{eq:mpc}, the requirement to have the MPC of CLDP less than or equal to that of LDP can be written as:

{\small
\begin{align} \label{eq:conv4}
\max_{v,y} &\frac{\pi(v) \cdot \text{Pr}[\Phi(v) = y]} {\sum_{z \in \mathcal{U}} \pi(z) \cdot \text{Pr}[\Phi(z) = y]} \leq \max_{v,y} \frac{\pi(v) \cdot \text{Pr}[\Psi(v) = y]} {\sum_{z \in \mathcal{U}} \pi(z) \cdot \text{Pr}[\Psi(z) = y]}
\end{align}
}
where $\Psi$ denotes LDP perturbation and $\Phi$ denotes CLDP perturbation. Using $\mathcal{U}$, $d$, $\pi$ and $\varepsilon$, we can compute the right hand side, and then search for the largest $\alpha$ such that the left hand side remains smaller than the right hand side, iteratively by incrementing $\alpha$ in each iteration and re-computing the left hand side in each iteration. The complexity of computing the right hand side (left hand side is analogous) is $\mathcal{O}(|\mathcal{U}|^3) \cdot \mathcal{O}(\Psi)$, where $\mathcal{O}(\Psi)$ is the complexity of the perturbation mechanism, often linear or sublinear in $\mathcal{U}$ or constant time. This is assuming access and querying of $\pi(v)$ and $d(v_i,v_j)$ are constant time operations, as they can be pre-computed in $\mathcal{O}(|\mathcal{U}| + |\mathcal{U}|^2)$ time and stored in a form that supports efficient access, which does not affect overall complexity. In deployment, the computation to convert $\varepsilon$ to $\alpha$ is a one-time cost performed at protocol setup time, therefore it does not have a negative impact on real-time user experience. 

An alternative to solving Equation \ref{eq:conv4} could be to obtain a closed-form relationship between $\alpha$ and $\varepsilon$ given $\mathcal{U}$, $d$, and $\pi$. We found that such a closed-form relationship can be established under particular cases such as when $\pi$ is uniform or when $d$ satisfies certain properties. However, the closed-form relationships require constructing and solving a high degree polynomial, which has higher execution time in practice than the iterative solution to Equation \ref{eq:conv4} which we presented above, due to the inherent hardness and time complexity of solving high degree polynomials. Hence, we give the current version of Equation \ref{eq:conv4} for wider practical applicability and more efficient implementation than the closed-form relationships we could derive. One of our ongoing future work directions is considering the derivation of efficient closed-form relationships.

\vspace{2pt}
\noindent\textbf{Roles of Relevant Factors:} Among the relevant factors, $\mathcal{U}$ impacts not only the outcome of $\varepsilon$ to $\alpha$ conversion, but also the computational complexity. This is because $\mathcal{U}$ is a relevant factor in the $\text{Pr}[\cdot]$ calculations in both LDP and CLDP. For example, the size of the bitvector in RAPPOR is affected by $|\mathcal{U}|$. EM of CLDP is also affected by $|\mathcal{U}|$, as each element in $\mathcal{U}$ must be assigned a score and the scores are then normalized. $d$ has no impact on the behavior of LDP, but it affects CLDP. Since our conversion aims to achieve equivalent protection in CLDP compared to LDP against the Bayesian adversary from Section \ref{sec:threatmodel}, it is expected that larger the $\max_{v_i, v_j}d(v_i,v_j)$, smaller the $\alpha$ value under the same $\varepsilon$. This can be explained via the fact that under fixed $\varepsilon$, according to the exponent in Definition \ref{def:CLDP}, when $d$ is larger, then $\alpha$ must be lowered to counter-balance the impact of increased $d$ to still achieve equivalent protection. $\pi$ is also a relevant factor in Equation \ref{eq:conv4}. As we will exemplify in the upcoming practical analysis, when $\pi$ is skewed (rather than uniform) it becomes the dominating factor in both the right and left hand sides of Equation \ref{eq:conv4}; therefore under the same $\varepsilon$, typically larger $\alpha$ is obtained when $\pi$ is skewed compared to uniform. Finally, the number of users does not appear in Equation \ref{eq:conv4} or play a role in the conversion, since the same $\varepsilon$ is used across all users in LDP and similarly the same $\alpha$ is used across all users in CLDP.

\begin{figure}[!t]
    \centering
    \includegraphics[width=.45\textwidth]{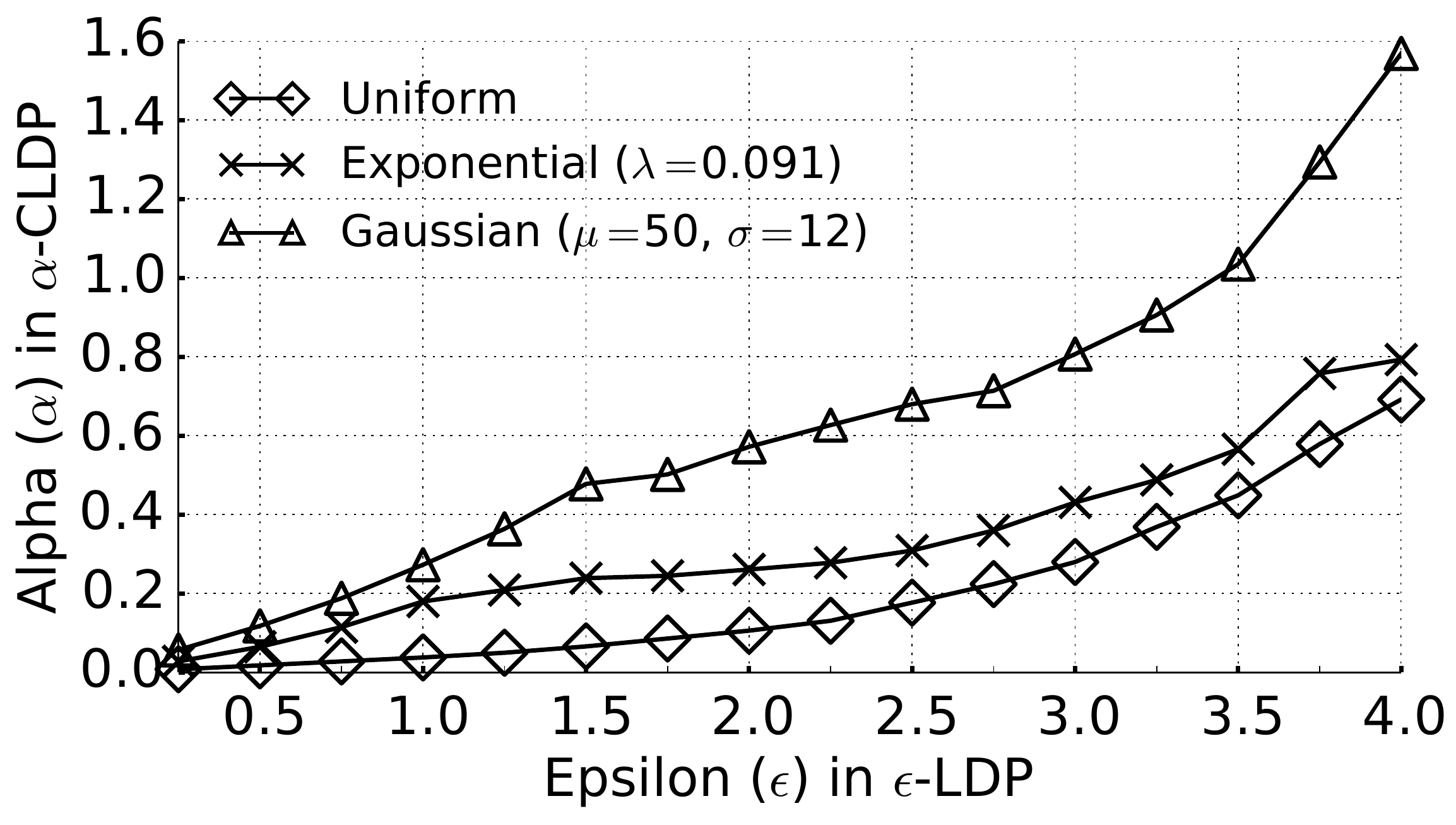}
    \vspace{-7pt}
    \caption{Exemplifying the relationship between $\varepsilon$ in $\varepsilon$-LDP and $\alpha$ in $\alpha$-CLDP according to Equation \ref{eq:conv4}, with $|\mathcal{U}|=100$ and $\pi=$ Uniform, Exponential, Gaussian.}
    \label{fig:conversion}
    \vspace{-11pt}
\end{figure}

\vspace{2pt}
\noindent\textbf{Practical Analysis:} To demonstrate the practicality of the relationship we establish above and derive insights, we solve Equation \ref{eq:conv4} under three example $\pi$ settings. In all settings, we assume $\mathcal{U}$ is the set of integers between $[0,99]$ and $d$ measures absolute value distance between two integers. We use $\pi$ = Uniform, Gaussian and Exponential distributions with the corresponding distribution parameters given in Figure \ref{fig:conversion}. Our rationale is that each $\pi$ represents a different type of skewness: Uniform has no skewness, Gaussian is symmetrically skewed around the mean, and Exponential has positive skew. Users' secrets are samples from these distributions rounded to the nearest integer. Distribution parameters are chosen so that users' secrets fall within the universe of $[0,99]$ with non-negligible tail probabilities. In Figure \ref{fig:conversion}, we provide the results of our study for a wide range of $\varepsilon$ values used in the literature: 0.25~$\leq \varepsilon \leq $~4. Note that this figure is obtained purely by solving Equation \ref{eq:conv4}, and does not require a real simulation or execution of the protocol involving end clients, i.e., the solution can be used in setup time before any data collection occurs. 

Figure \ref{fig:conversion} allows us to derive several interesting insights. The first important conclusion is that $\varepsilon$ and $\alpha$ are positively correlated -- as the privacy requirement of LDP is relaxed, we can also relax that of CLDP. Second, it is often the case that LDP and CLDP give equivalent protection when $\alpha \ll \varepsilon$ holds. This is because in order to compensate for the added term of $d(\cdot,\cdot)$ decreasing indistinguishability of distant items under CLDP, we have to use much smaller $\alpha$ in CLDP compared to $\varepsilon$ of LDP. Third, we observe that the relationship between $\varepsilon$ and $\alpha$ depends on $\pi$, e.g., if the data follows a Uniform distribution, CLDP must use a stricter $\alpha$ than the other two distributions. The reason is because there are two determining factors in calculating adversarial confidence: $\pi$ and $\text{Pr}[f(v)=y]$. For skewed distributions, $\pi$ becomes the dominating factor and since the same $\pi$ is shared by LDP and CLDP, the behavior of perturbation functions have relatively less impact on adversarial confidence. In contrast, for the Uniform distribution, since $\text{Pr}[f(v)=y]$ becomes the dominating factor and its value is high for the tail-ends of the domain in the case of CLDP, we must use lower (stricter) $\alpha$ in CLDP to match the adversarial confidence in LDP. Based on our findings and observation that the Uniform $\pi$ setting causes lowest $\alpha$ under the same $\varepsilon$ compared to other $\pi$ settings, in the remainder of the paper (including our experiments in Section \ref{sec:experiments}) we assume the Uniform $\pi$ setting. This helps demonstrate CLDP's utility benefit in the strictest and most challenging setting.

\vspace{2pt}
\noindent\textbf{Note on Alternate Threat and Adversary Models:} The analysis we perform in this section and the rest of the paper is under the Bayesian adversary and measurement of MPC described in Section \ref{sec:threatmodel}. Since Bayesian methods were previously used in the literature for evaluating DP variants and relaxations in different domains, we argue that the Bayesian adversary model is relevant and representative \cite{li2013membership,yang2015bayesian,kasiviswanathan2014semantics,yu2017dynamic}. Yet, we acknowledge that under different adversaries, the protection offered by LDP and CLDP could be different, and thus, a different conversion between $\varepsilon$ and $\alpha$ could be needed. For example, one alternative method to compare LDP and CLDP can be hypothesis testing \cite{ding2018detecting,kairouz2017composition,liu2019investigating}. We leave comparison between LDP and CLDP under such different protection models to future work, since the main contribution of this paper is creating practical CLDP protocols and demonstrating their utility benefit under the Bayesian adversary model. 

\vspace{-5pt}
\section{CLDP Mechanisms and Protocols}

In this section, we present protocols that can be used in practice to collect data while achieving CLDP. We present three protocols: Ordinal-CLDP, Item-CLDP, and Sequence-CLDP, to address different types of client data.

\vspace{-4pt}
\subsection{Ordinal-CLDP for Ordinal Items}

Our first protocol is Ordinal-CLDP, which addresses data types that stem from finite metric spaces, i.e., $\mathcal{U}$ is discrete and finite, and there exists a built-in distance metric $d: \mathcal{U} \times \mathcal{U} \rightarrow [0,\infty)$. This setting covers a variety of useful data types: (i) discrete numeric or integer domains where $d$ can be the absolute value distance between two items, (ii) ordinal item domains with total order, e.g., letters and strings ordered by dictionary order A $<$ B $<$ C $<$ ..., and (iii) categorical domains with tree-structured domain taxonomy where distance between two items can be measured using the depth of their most recent common ancestor in the taxonomy tree \cite{soria2014enhancing,sanchez2012ontology}. In these scenarios, item order and $d$ are naturally defined and enforced.

In Ordinal-CLDP, each client locally applies the Exponential Mechanism (EM) implementation shown in Algorithm \ref{alg:EM}, and uploads the perturbed output to the server. Notice that $\alpha$, $\mathcal{U}$, $d$, and user's true value $v$ are all inputs to the algorithm. The algorithm's output $v'$ is sent to the collector, thereby concluding the protocol in a single round without blocking.

\setlength{\textfloatsep}{6pt}
\begin{algorithm}[!t]
\caption{Ordinal-CLDP using EM}
\DontPrintSemicolon
\label{alg:EM}
\SetKwInOut{Input}{Input}\SetKwInOut{Output}{Output}
\Input{$\alpha$: CLDP privacy parameter,~~ $\mathcal{U}$: item universe, \\ $d$: distance metric,~~ $v$: user's true value}
\Output{$v' \in \mathcal{U}$: perturbed value}
\BlankLine
\For{each $y \in \mathcal{U}$}{
Assign $\text{score}(y) = e^{\frac{-\alpha \cdot d(v,y)}{2}}$ \;
}
Pick a random sample $v'$ from $\mathcal{U}$, where $\text{Pr}[v' \text{ is sampled}] = \frac{\text{score}(v')}{\sum_{z \in \mathcal{U}} \text{score}(z)}$ \;
\textbf{return} $v'$ \;
\end{algorithm} 

Algorithm \ref{alg:EM} has some desirable utility properties. First, since $d$ is a metric, it holds that $d(v,v)=0$ for all $v \in \mathcal{U}$. As a result, each invocation of Algorithm \ref{alg:EM} is unbiased since $\text{Pr}[\Phi_{ORD}(v)=v] > \text{Pr}[\Phi_{ORD}(v)=y]$ where $y \neq v$ and $\Phi_{ORD}$ denotes Algorithm \ref{alg:EM}. Furthermore, when $d$ is selected with the property that $d(v_i,v_j)=c$ for all $v_i, v_j \in \mathcal{U}$ where $v_i \neq v_j$ and $c$ is a constant such as $c=1$, then the aggregation of the outputs of Algorithm \ref{alg:EM} on the server side preserves the relative frequency relationship between $v_i$ and $v_j$ in expectation. In other words, if the true frequency of $v_i$ is larger than $v_j$ [resp. smaller than], after each client uploads perturbed items resulting from Algorithm \ref{alg:EM} and the counts of the perturbed items are computed on the server side, the observed frequency of $v_i$ is expected to be larger than the observed frequency of $v_j$ [resp. smaller]. Given that frequency relationships between individual pairs of items are preserved, general item frequency rankings are also preserved. We provide the proof of this property in the appendix.

\subsection{Item-CLDP for Non-Ordinal Items} \label{sec:itemCLDP}

Our second protocol is Item-CLDP in which each user still holds a singleton true item, but the items come from an arbitrary $\mathcal{U}$ with no pre-defined $d$ or total order. For example, if $\mathcal{U}$ consists of OS names, an order of MacOS $<$ Ubuntu $<$ Windows is neither available nor initially justifiable. This non-ordinal item setting has been assumed in recent LDP research for finding popular emojis, emerging slang terms, popular and anomalous browser homepages, and merchant transactions \cite{thakurta2017emoji,thakurta2017learning,erlingsson2014rappor,wang2017locally,bassily2017practical}. We propose Item-CLDP in a generic way to maximize its scope and cover such existing cases. Parallel to previous works, our goal is to uphold relative item frequencies to learn popularity histograms and discover heavy hitters. To this end, we propose that a desirable perturbation strategy should replace a \textit{popular} item with another popular item, and an \textit{uncommon} item with another uncommon item. This achieves our goal of upholding relative item frequencies, as the expected behavior (conceptually) will be that popular items and uncommon items will be shuffled among themselves, and relative frequencies will be preserved.

\begin{figure}[!t]
    \centering
    \includegraphics[width=.49\textwidth]{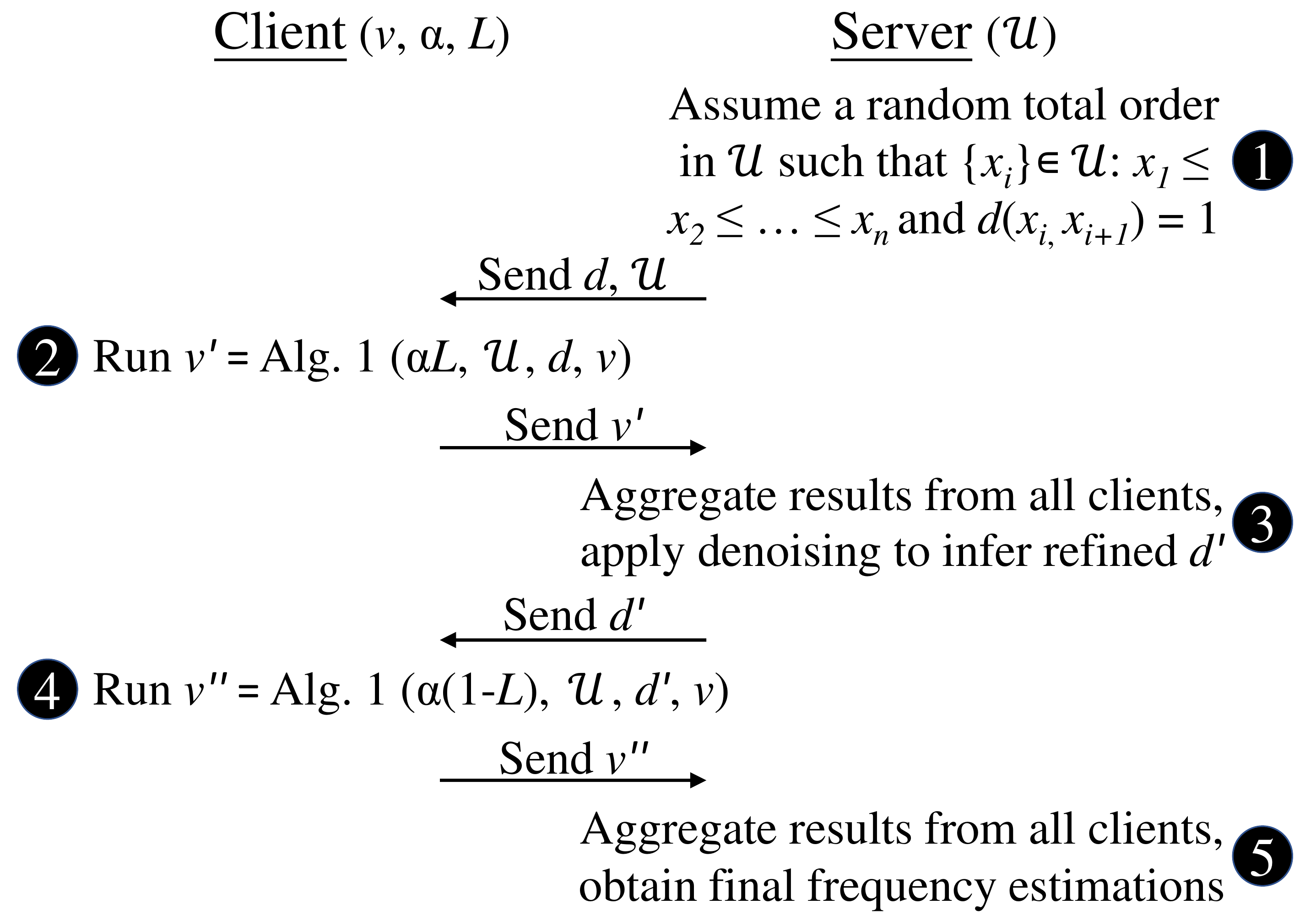}
    \vspace{-10pt}
    \caption{Item-CLDP protocol to report non-ordinal items.} 
    \label{fig:itemCLDP}
    \vspace{4pt}
\end{figure}

The proposed Item-CLDP protocol is given in Figure \ref{fig:itemCLDP}. In Item-CLDP the server communicates with each client twice, hence the protocol consists of two rounds. The first round contains steps 1-3 and the second round contains steps 3-5. The server executes the first round with each client in parallel (without blocking). At the end of the first round, the server performs the aggregation and de-noising step (Step 3). Then, the server executes the second round of communication. Next, we explain each step in detail. 

\vspace{2pt}
\noindent\textbf{Step 1.} When the protocol starts, the server knows universe $\mathcal{U}$ and each client has a true value $v$. The value of the privacy budget $\alpha$ and budget allocation parameter 0~$< L <$~1 can be publicly known. (The role of $L$ will be explained later.) A random total order is constructed among all items in $\mathcal{U}$ such that for $\{x_i\} \in \mathcal{U}$, $d(x_i, x_{i+1})=1$. The server advertises $\mathcal{U}$ and $d$ to all clients.

\vspace{2pt}
\noindent\textbf{Step 2.} Each client runs Algorithm \ref{alg:EM} with budget $\alpha \cdot L$ to obtain a perturbed value $v'$ locally on their device. Then, the clients send their $v'$ to the server.

\vspace{2pt}
\noindent\textbf{Step 3.} Due to the utility-unaware choice of $d$ in Step 1, the absolute item frequencies discovered at this step contain significant error. A second round is desirable to reduce error. We found that although the server does not accurately learn absolute item frequencies by the end of Step 2, it can learn frequency \textit{ranking} of items after applying a de-noising strategy. A key aspect is how de-noising is performed. Let $y \in \mathcal{U}$ be an item, let $\text{true}(y)$ be the true count of $y$ in the population, which we are trying to find, and let $\text{obs}(y)$ denote the observed count of $y$ following the first round of client-server communication (i.e., by the beginning of Step 3). The following holds in expectation:
\begin{align*}
\text{obs}(y) =~ &\text{true}(y) \cdot \text{Pr}[\Phi_{EM}(y) = y] \\&+ \sum_{x \in \mathcal{U} \setminus \{y\}} \text{true}(x) \cdot \text{Pr}[\Phi_{EM}(x) = y] 
\end{align*}

Our goal is to solve for $\text{true}(y)$, but we cannot do so since $\text{true}(x)$ is also unknown. Hence, in our de-noising strategy we make the heuristic decision of plugging $\text{obs}(x)$ in place of $\text{true}(x)$, thereby obtaining $\text{true}'(y)$ as follows:
\[
\text{true}'(y) = \frac{\text{obs}(y) - \sum_{x \in \mathcal{U} \setminus \{y\}} \text{obs}(x) \cdot \text{Pr}[\Phi_{EM}(x) = y]} {\text{Pr}[\Phi_{EM}(y) = y]}
\]
We apply this to all $y$'s in $\mathcal{U}$ and rank them according to their $\text{true}'(y)$. The distance function $d'$ is set to reflect this new ranking instead of the original $d$ from Step 1.

\vspace{2pt}
\noindent\textbf{Step 4.} After the clients receive $d'$, each client runs Algorithm~\ref{alg:EM} with $d'$ to obtain $v''$ and sends $v''$ to the server. This invocation of the algorithm is with budget $(1-L) \cdot \alpha$. 

\vspace{2pt}
\noindent\textbf{Step 5.} Upon receiving the $v''$ values from all clients, the server aggregates all results and obtains the final frequency estimates.

\vspace{2pt}

Since Item-CLDP is a two-round protocol, each client sends perturbed information twice. Hence, we need to quantify the total disclosure by the end of two rounds. We introduce the parameter $L$, which takes values between 0 and 1 and determines how the CLDP privacy budget will be allocated to the two rounds of Item-CLDP. Denoting Item-CLDP by $\Phi_{ITEM}$, we can show that for any two possible inputs $v_1$,$v_2$ of a user:
\begin{align*}
\frac{\text{Pr}[\Phi_{ITEM}(v_1) = \langle v', v'' \rangle]}{\text{Pr}[\Phi_{ITEM}(v_2) = \langle v', v'' \rangle ]} &\leq e^{\alpha \cdot L \cdot d(v_1,v_2)} \cdot e^{\alpha \cdot (1-L) \cdot d'(v_1,v_2)} \\
&\leq e^{\alpha \cdot \max\{d(v_1,v_2), d'(v_1,v_2)\}}
\end{align*}
This follows from the fact that the first round satisfies $(\alpha \cdot L)$-CLDP with $d$, and the second round satisfies $\alpha \cdot (1-L)$-CLDP with $d'$. We choose the value of $L$ by finding which $L$ yields minimum frequency estimation error by the end of the second round of Item-CLDP. According to our experiments with different $L$, we recommend $L \cong 0.8$ as it often gives best results, which indicates that finding an accurate preliminary ranking in the first round of Item-CLDP is indeed important to obtain a good final result. An important property enabling the composition of Item-CLDP in above fashion is the fact that by construction of $d$ and $d'$ as metrics, $\max\{d(v_1,v_2), d'(v_1,v_2)\}$ is also a metric. This enables composition of the two individual CLDP rounds, and ensures their composed outcome achieves CLDP. The proof of composition is provided in the appendix.

\setlength{\textfloatsep}{\textfloatsepsave}

\vspace{-4pt}
\subsection{Sequence-CLDP for Item Sequences}

In Ordinal-CLDP and Item-CLDP, each user reports a single item. We now study the case where each user reports a collection of items. We give our Sequence-CLDP protocol assuming this collection forms a sequence and later show the applicability of Sequence-CLDP to set-valued data. Sequential data arises naturally in many domains, including cybersecurity (log files), genomics (DNA sequences), web browsing histories, and mobility traces; thus, a protocol for privacy-preserving collection of item sequences holds great practical value. We denote by $X$ a user's true sequence, and by $X[i]$ the $i$'th element in $X$. Each element $X[i]$ is an item from universe $\mathcal{U}$. We assume the distance metric $d$ between individual items is known apriori, e.g., for ordinal $\mathcal{U}$ we can use built-in $d$ as in Ordinal-CLDP; otherwise, we can infer $d$ using a process similar to the first round of Item-CLDP. We measure distance between two sequences $d_{\text{seq}}(X,Y)$ as:
\vspace{-3pt}
\[
d_{\text{seq}}(X,Y) = \sum_{i=1}^{|X|} d(X[i],Y[i])
\]
\vspace{-5pt}

In Sequence-CLDP, each client runs the sequence random\-ization procedure given in Algorithm \ref{alg:sequenceCLDP} to locally perturb their $X$. The procedure has two probability parameters: 0 $<$ \textit{halt, gen} $<$ 1, and a length parameter \textit{max\_len} denoting the maximum sequence length allowed. Given true sequence $X$, the algorithm returns a perturbed sequence $S$. Our goal in Sequence-CLDP is to hide two complementary types of information: the \textit{length} of $X$ and the \textit{contents} of $X$. For example, let $X$ consist of a sequence of security events observed on a machine. Hiding the length of $X$ is useful because it disables the adversary from learning that many security events were observed on this machine, hence that the machine is probably infected. Hiding the contents of $X$ is useful because it disables the adversary from learning \textit{which} security events were observed, hence the adversary cannot infer which types of problems exist on the machine, which attacks are successful, and so forth. Denoting Sequence-CLDP by $\Phi_{SEQ}$, we formalize these privacy properties as follows.

\begin{definition} \label{def:length}
Let $\text{Pr}[\Phi_{SEQ}(X) \leadsto \ell]$ denote the probability that $\Phi_{SEQ}(X)$ produces a perturbed sequence of length $\ell$ given input sequence $X$. We say that $\Phi_{SEQ}$ satisfies $\alpha$-length-indistinguishability if for any pair of true sequences $X$, $Y$:
\[
\frac{\text{Pr}[\Phi_{SEQ}(X) \leadsto \ell]} {\text{Pr}[\Phi_{SEQ}(Y) \leadsto \ell]} \leq e^{\alpha \cdot abs(|X|-|Y|)}
\]
\end{definition}

\begin{definition} \label{def:content}
Let $\text{Pr}[\Phi_{SEQ}(X) = S]$ denote the probability that $\Phi_{SEQ}(X)$ produces perturbed sequence $S$ given input sequence $X$. We say that $\Phi_{SEQ}$ satisfies $\alpha$-content-indistinguishability if, for any pair of true sequences $X$, $Y$ of same length, it holds that:
\[
\frac{\text{Pr}[\Phi_{SEQ}(X) = S]} {\text{Pr}[\Phi_{SEQ}(Y) = S]} \leq e^{\alpha \cdot d_{\text{seq}}(X,Y)}
\]
\end{definition}

It can be observed that both properties are adaptations of CLDP for hiding the two types of sensitive information relating to sequences: length and content. Definition \ref{def:length} is analogous to CLDP with metric $d$ being the absolute value difference between sequence lengths. It ensures that an adversary observing the length of the perturbed sequence $\ell$ cannot infer the length of the user's true sequence with high confidence. Definition \ref{def:content} is analogous to CLDP with metric $d$ being $d_{\text{seq}}$, i.e., item-wise difference in sequences' contents. It ensures that an adversary observing the contents of the perturbed sequence $S$ cannot infer the contents of the user's true sequence with high confidence. The combined privacy protections offered by Definition \ref{def:length} and Definition \ref{def:content} can be best explained from the following perspective: Given that the adversary observes a perturbed sequence $S$ with length $\ell = |S|$ that a user sends to the data collector, can the adversary reverse-engineer the length or the contents of the user's true sequence? If the perturbation satisfies Definitions \ref{def:length} and \ref{def:content} simultaneously, then the answer is negative, since both sequence length and content are protected analogously to the protection offered by CLDP.

\begin{theorem} \label{thm:sequenceCLDP}
Algorithm \ref{alg:sequenceCLDP} satisfies $\alpha$-length-indistin\-guishability and $\alpha$-content-indistinguishability simultaneously if \textit{halt}, \textit{gen} are selected either symmetrically as:
\[
\textit{halt} = \textit{gen} = \frac{1}{e^{\alpha}+1}
\]
or asymmetrically within the ranges:
\[
0 < \textit{halt} < \frac{1}{e^{\alpha}+1} ~~\text{and}~~ 1 - e^{\alpha} \cdot \textit{halt} \leq \textit{gen} \leq 1 - \frac{\textit{halt}}{e^{\alpha}}
\]
\end{theorem}
\vspace{-6pt}
\begin{proof}
Provided in the appendix.
\end{proof}

\setlength{\textfloatsep}{6pt}
\begin{algorithm}[t]
\caption{Sequence-CLDP randomization}
\DontPrintSemicolon
\label{alg:sequenceCLDP}
\SetKwInOut{Input}{Input}\SetKwInOut{Output}{Output}\SetKwInOut{Parameters}{Parameter}
\Parameters{Probabilities \textit{halt}, \textit{gen}, maximum length allowed \textit{max\_len}}
\Input{Client's private sequence $X$, privacy budget $\alpha$, item universe $\mathcal{U}$, distance metric $d$} 
\Output{Randomized sequence $S$}
\BlankLine
\While{$X$ is shorter than \textit{max\_len}} {
Pad $X$ with stop sign $\bot$ \tcp*{Dummy symbol} 
} 
Initialize empty sequence $S$ \;
\For{$i=1$ to \textit{max\_len}} {
\If{$X[i] \neq \bot$} {  
With probability \textit{halt}, stop here and return $S$\;
With probability $(1-\textit{halt})$, run Alg.~\ref{alg:EM} with inputs ($\alpha$, $\mathcal{U}$, $d$, $X[i]$) to obtain $v'$, and append $v'$ to $S$
} \Else{ 
With probability \textit{gen}, randomly pick an item from $\mathcal{U}$ and append it to $S$\;
With probability $(1-\textit{gen})$, stop here and return $S$\;
}
}
\textbf{return} $S$ \;
\end{algorithm}

In Algorithm \ref{alg:sequenceCLDP}, high \textit{halt} causes the algorithm to terminate early for a long sequence, causing the perturbed sequence $S$ to be much shorter than $X$. High \textit{gen} adds random items to $S$, thus it causes $S$ to be much longer than $X$; in addition, since the added items are sampled uniformly at random, $S$ will contain bogus elements. If we consider only the utility perspective, simultaneously decreasing the values of \textit{halt} and \textit{gen} yields higher sequence utility. However, Theorem \ref{thm:sequenceCLDP} places bounds on the values of \textit{halt} and \textit{gen}; we cannot arbitrarily decrease them, otherwise we will not satisfy the indistinguishability properties. Among the given choices, asymmetric parameter choice is preferable when we expect users' true sequences to be long, since it assigns a lower halting probability compared to the symmetric case, thereby decreasing the probability that the algorithm is terminated early. This is done at the cost of increased \textit{gen}, which implies that the asymmetric case will more likely add synthetic elements to $S$. Since this is detrimental to utility especially when users' true sequences are short, for short sequences, we recommend using the symmetric parameter choice.

\vspace{2pt}
\noindent\textbf{Application to set-valued data.} Although Sequence-CLDP is designed for sequences, it can be applied to set-valued data without information loss as follows. First, each user enforces a random ordering among the items in their itemset, to convert the itemset to a sequence. Second, the user runs Sequence-CLDP on this converted sequence to obtain a perturbed sequence. Third, the user removes the ordering from the perturbed sequence to obtain a perturbed itemset. Finally, the perturbed itemset is sent to the server. On the other hand, we cannot use existing set-valued LDP protocols \cite{qin2016heavy,wang2018locally} on sequences without losing their sequentiality (ordering) aspect. Hence, we believe Sequence-CLDP has wider applicability than existing set-valued protocols.

\section{Experimental Evaluation} \label{sec:experiments}

We compare our proposed CLDP protocols against the existing LDP protocols on real-world cybersecurity datasets provided by Symantec as well as on public datasets. In singleton item comparison, we use RAPPOR (proposed and deployed by Google \cite{erlingsson2014rappor}) and OLH (recent protocol with improved utility over prior works \cite{wang2017locally}). In set-valued setting, we use SVIM as the current state-of-the-art LDP protocol \cite{wang2018locally}. In each experimental dataset and setting, given $\mathcal{U}$, $d$ and $\varepsilon$, we freshly execute Equation \ref{eq:conv4} and the process in Section \ref{sec:conversion} to obtain the appropriate $\alpha$ parameter for CLDP to ensure a fair comparison under each individual setting. Uniform $\pi$ is used in each execution of Equation \ref{eq:conv4} since it is the most challenging $\pi$ setting for CLDP to demonstrate its utility.

To ensure our experimental comparison between LDP and CLDP is fair with respect to the level of privacy protection under the Bayesian adversary model, we perform the following steps. First, we simulate data collection with LDP protocols and chosen $\varepsilon$, and measure the utility loss and privacy protection in terms of MPC. Second, we use the technique in Section \ref{sec:conversion} to determine the $\alpha$ that should be used for CLDP to match the protection achieved by LDP. Third, we simulate data collection with CLDP protocols and $\alpha$ from the previous step. Finally, we compare the utility loss caused by CLDP versus the utility loss of LDP.

In Section \ref{sec:cybersecurity_datasets}, we consider cybersecurity use cases that reflect the limitations of existing LDP protocols, e.g., user populations are small and sequential datasets cannot be handled. In these experiments, our results show that LDP protocols do not yield sufficient utility while our CLDP protocols offer satisfactory utility in most cases, hence their use in corresponding security products is practical and preferable. In Section \ref{sec:public_datasets}, we experiment on public datasets which differ from the above since they do not reflect the limitations of LDP protocols, e.g., user populations are sufficiently large (over half a million). Even so, we show that our CLDP protocols perform at least comparable to, or in many cases better than, existing LDP protocols. An interesting result is that while LDP protocols are capable of finding the frequencies and ranking of heaviest hitter items with good accuracy (e.g., top 5-10\% of the universe), CLDP protocols' accuracy is similar for these few heavy hitters, but they significantly outperform LDP protocols for remaining items (medium frequency and infrequent items).

\setlength{\textfloatsep}{\textfloatsepsave}

\begin{figure*}[!ht]
\centering
\includegraphics[width=0.93\textwidth]{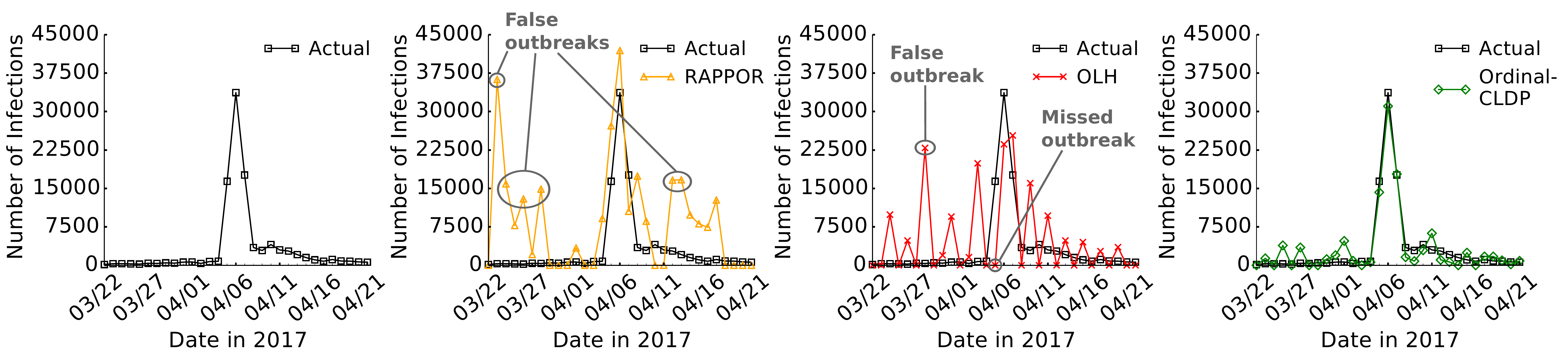} 
\vspace{-8pt}
\caption{Monitoring the infections of ransomware Cerber for one month to detect a potential outbreak. Illustrated in the graphs is the actual number of infections versus infections reported by LDP protocols RAPPOR and OLH, and our Ordinal-CLDP protocol ($\varepsilon$~=~1). Ordinal-CLDP enables accurate recovery of daily infection counts and detection of outbreaks without major false positives or negatives.} 
\label{fig:cerber}
\vspace{-10pt}
\end{figure*}

\begin{figure*}[!ht]
\centering
\includegraphics[width=0.93\textwidth]{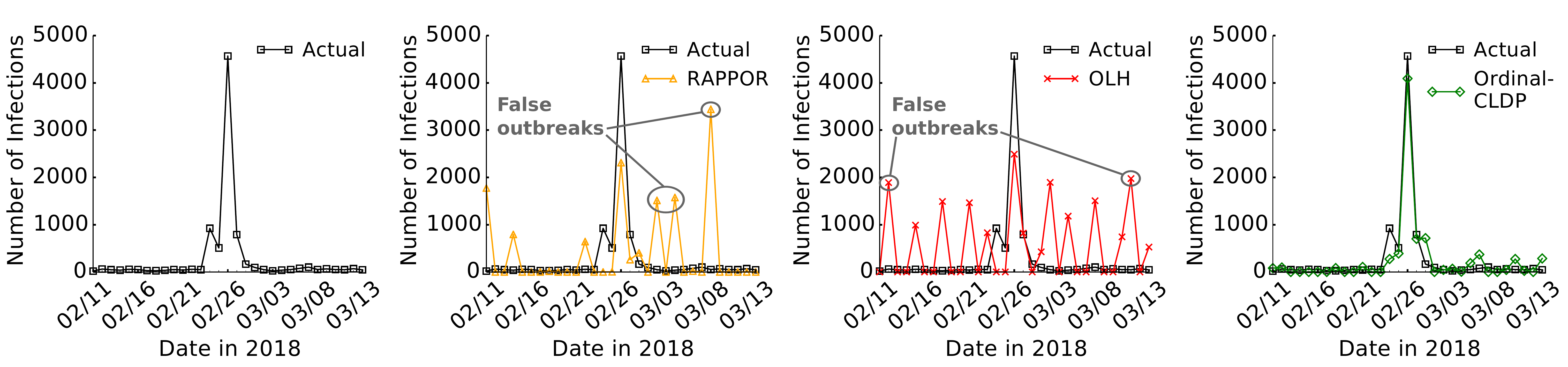} 
\vspace{-8pt}
\caption{Monitoring the infections of ransomware Locky for one month to detect a potential outbreak. Refer to Figure~\ref{fig:cerber} for a similar experiment with a different ransomware ($\varepsilon$~=~1.0). Same conclusions apply to this figure.}
\label{fig:locky}
\vspace{-8pt}
\end{figure*}

\vspace{-5pt}
\subsection{Case Studies with Cybersecurity Datasets} \label{sec:cybersecurity_datasets}

Cybersecurity datasets provided by Symantec allowed us to test the accuracy of our protocols on pertinent real-world use cases and assess their practical applicability. We note that certain details such as the total number of infected machines are omitted on purpose due to confidentiality reasons.

\vspace{2pt}
\noindent
\underline{\textbf{Case Study \#1: Ransomware Outbreak Detection}}
\vspace{1pt}

\textbf{Setup.} We consider the case where Symantec collects malware reports from machines running its anti-malware protection software. Each machine sends a locally private malware report to Symantec daily, containing the count of malware-related events observed on that machine during that day. Privacy is injected into malware reports by modifying the actual counts with LDP/CLDP.

For our experiments, we obtained the daily infection counts of two ransomware variants within those time periods in which we already know there were global outbreaks. Specifically, we considered the infections reported for \textit{Cerber} between March 22 and April 21 in 2017 with the outbreak happening on April 6, and those reported for \textit{Locky} between February 11 and March 13 in 2018 with the outbreak happening on February 26. We evaluated how accurately the total number of daily infections for these two ransomware variants can be estimated using our Ordinal-CLDP approach versus LDP approaches RAPPOR and OLH. The goal of our experiment is to retroactively test whether LDP/CLDP could identify if and when a ransomware outbreak happened.

\textbf{Results.} 
We illustrate the results in Figures \ref{fig:cerber} and \ref{fig:locky} for Cerber and Locky respectively. If we use RAPPOR or OLH to perform detection, we obtain many false positives (days on which RAPPOR/OLH claim there was an outbreak, but in fact there was not) and false negatives (days on which RAPPOR/OLH claim there was no outbreak, but in fact there was). 
Some important examples are marked on the graphs, e.g., in Figure \ref{fig:cerber}, RAPPOR raises false positives on March 23-24 as well as April 12-13. In addition, OLH misses the onset of the outbreak happening on April 5 by reporting 0 observed infections whereas in reality there are 16,388 infections. False positives are costly to Symantec since they cause the company to devote resources and response teams to combat a malware outbreak that does not exist. False negatives are also costly since Symantec will not react to an outbreak in a timely manner, losing customer trust. Observing this many false positives and false negatives with LDP methods raises serious concerns. In contrast, using Ordinal-CLDP, Symantec can obtain daily infection counts with high accuracy. Note that in Figures \ref{fig:cerber} and \ref{fig:locky}, there are small discrepancies between the actual infections versus CLDP's predicted infections, which demonstrates that CLDP is not error-free. Nevertheless, using CLDP ransomware outbreaks can be detected in a privacy-preserving manner without major false positives or negatives.

\vspace{2pt}
\noindent
\underline{\textbf{Case Study \#2: Ransomware Vulnerability Analysis}}
\vspace{1pt}

\textbf{Setup.} Next, we ask the question: Can we find which operating systems were most infected by ransomware? This would assist Symantec in discovering vulnerable or targeted OSs. When performing this analysis, we focus specifically on the day of outbreak (April 6, 2017 for Cerber and February 26, 2018 for Locky) and the machines reporting infections on this day. We assume Symantec obtains a locally private malware report from these machines including the vendor, specs, and OS version. Upon collecting reports from all machines, Symantec infers how frequently each OS was infected in the population, and ranks OSs in terms of infection frequency.

We conduct two experiments. In the first experiment, we find the actual (non-private) infection frequency of each OS, and compare actual frequencies with the frequencies that would be obtained if RAPPOR, OLH, or Item-CLDP were applied, using L1 distance as measurement of error. We vary $\varepsilon$ between 0.5~$\leq \varepsilon \leq $~4. In the second experiment, we fix $\varepsilon$~=~1, rank the OSs in terms of infection frequency (highest to lowest), and study the top-10 most infected OSs. Due to ethical considerations, we anonymize OS names by renaming according to their actual rank, e.g., top-ranked OS is named os1, 2nd ranked OS is named os2, and so forth. If a lower-ranked OS is a different version of a higher ranked OS, we add the version information to the name, e.g., os1 v2.

\begin{figure}
  \begin{minipage}[!b]{0.49\linewidth}
    \centering
    \includegraphics[width=\linewidth]{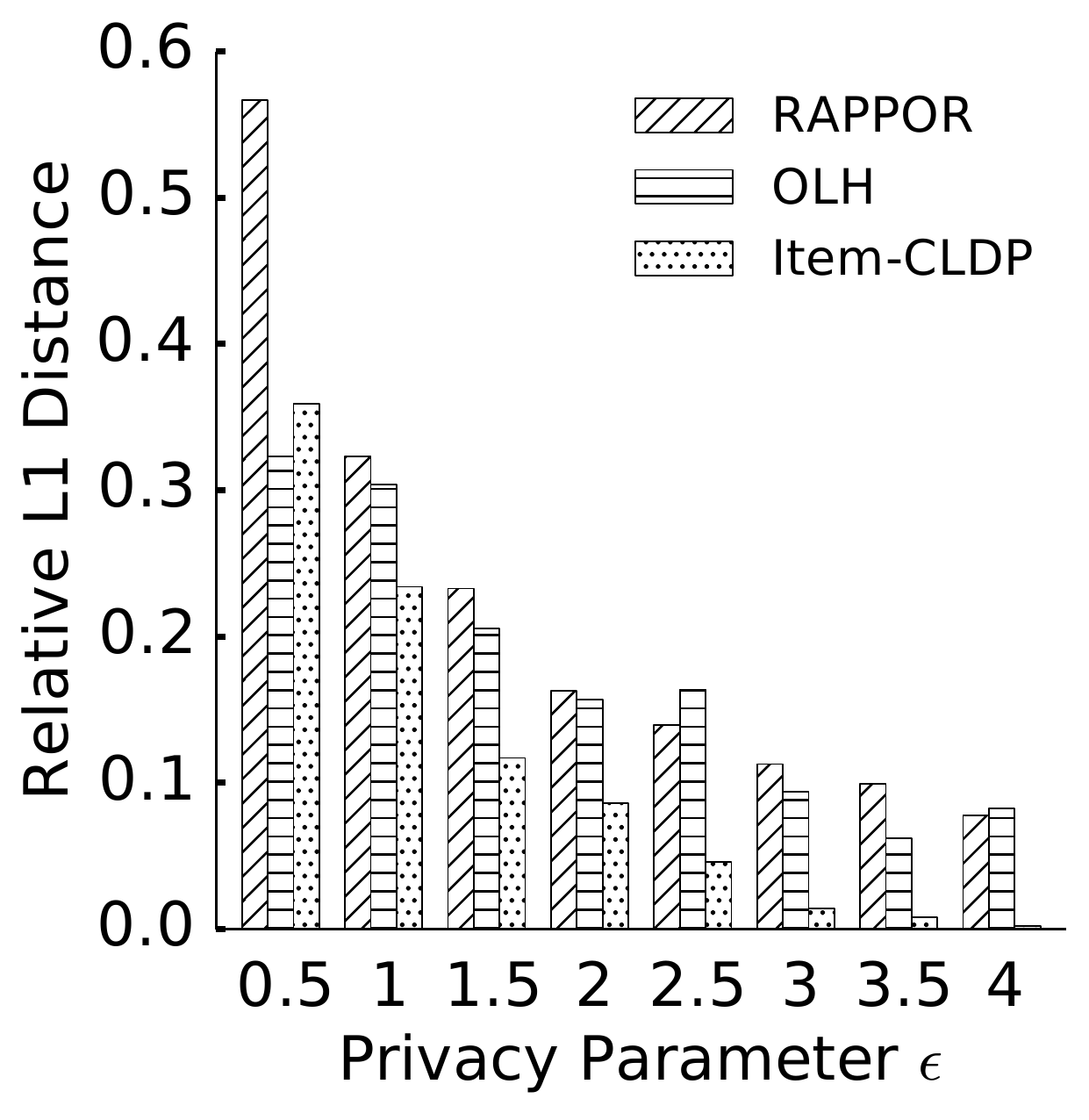}
  \end{minipage}%
  \begin{minipage}[!b]{0.49\linewidth}
    \centering
\begin{adjustbox}{width=0.99\columnwidth,totalheight=\textheight,keepaspectratio}
\setlength{\tabcolsep}{2.5pt}
\begin{tabular}{r | l | l | l | l}
\toprule
 & Actual & \shortstack[l]{RAP-\\POR} & OLH & \shortstack[l]{Item-\\CLDP}\\
\midrule
1 & \textbf{os1} & \textbf{os1} & \textbf{os1} & \textbf{os1}\\
2 & \textbf{os2} & \textbf{os2} & \textbf{os2} & \textbf{os2}\\
3 & \textbf{os2 v2} & \textbf{os2 v2} & \textbf{os2 v2} & \textbf{os2 v2}\\
4 & \textbf{os3} & \st{os2 v5} & \st{os2 v6} & \textbf{os3}\\	
5 & \textbf{os3 v2} & \st{os2 v6} & os1 v2 & os4\\
6 & \textbf{os4} & os3 & \st{os8} & os5\\
7 & \textbf{os5} & \st{os2 v7} & \st{os6 v3} & os3 v2\\
8 & \textbf{os2 v3} & \st{os6} & \st{os2 v8} & \textbf{os2 v3}\\
9 & \textbf{os2 v4} & \st{os7} & \st{os2 v5} & \textbf{os2 v4}\\
10 & \textbf{os1 v2} & \st{os6 v2} & os3 & \st{os8}\\
\bottomrule
\end{tabular}
\end{adjustbox}
\end{minipage}
\vspace{-2pt}
\caption{
Analyzing OS vulnerability for ransomware Cerber. 
\textit{Left}: L1 distances between actual OS counts versus locally private counts reported by RAPPOR, OLH, and our Item-CLDP protocol.
\textit{Right}: Actual and locally private top-10 OS rankings when $\varepsilon$~=~1. \textbf{Bold} = OS with correct rank, regular font = OS with wrong rank, \st{strike-through} = OS not in actual top-10.
Item-CLDP better preserves relative rankings and generally has lower L1 errors.} 
\label{fig:cerberOS}
\vspace{-5pt}
\end{figure}

\begin{figure}
  \begin{minipage}[!b]{0.49\linewidth}
    \centering
    \includegraphics[width=\linewidth]{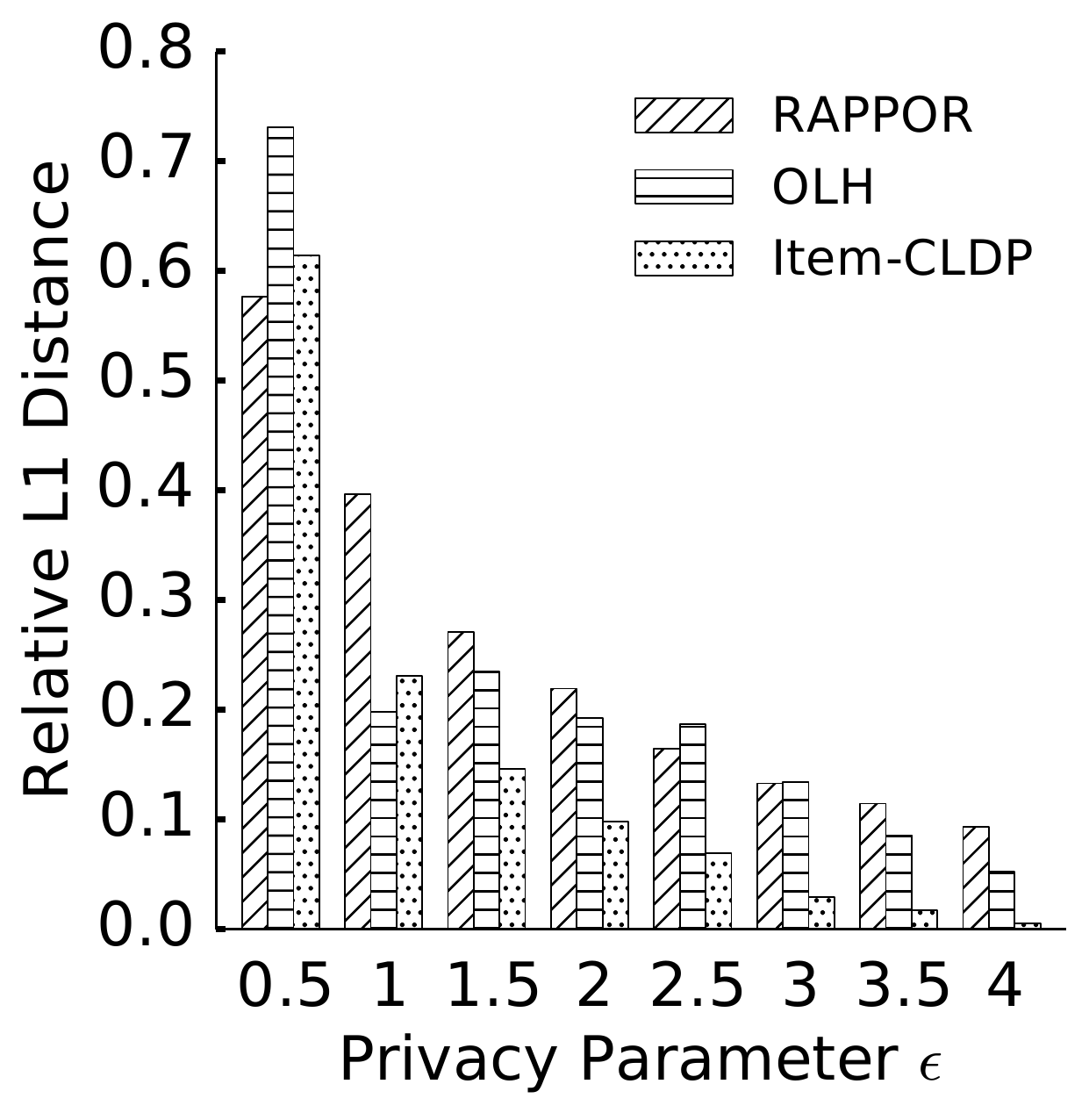}
  \end{minipage}%
  \begin{minipage}[!b]{0.49\linewidth}
    \centering
\begin{adjustbox}{width=0.99\columnwidth,totalheight=\textheight,keepaspectratio}
\setlength{\tabcolsep}{2.5pt}
\begin{tabular}{r | l | l | l | l}
\toprule
& Actual & \shortstack[l]{RAP-\\POR} & OLH & \shortstack[l]{Item-\\CLDP}\\
\midrule
1 & \textbf{os1} & \textbf{os1} & \textbf{os1} & \textbf{os1}\\
2 & \textbf{os2} & \textbf{os2} & \textbf{os2} & \textbf{os2}\\
3 & \textbf{os1 v2} & \st{os5 v2} & os3 & \textbf{os1 v2}\\
4 & \textbf{os1 v3} & os1 v5 & os4 & \st{os5 v3}\\	
5 & \textbf{os3} & os1 v3 & os2 v2 & \textbf{os3}\\
6 & \textbf{os1 v4} & os1 v2 & \textbf{os1 v4} & os5\\
7 & \textbf{os4} & os3 & \textbf{os4} & \st{os6}\\
8 & \textbf{os1 v5} & os5 & \st{os8} & \textbf{os1 v5}\\
9 & \textbf{os5} & \st{os6} & os1 v2 & \st{os7}\\
10 & \textbf{os2 v2} & \st{os7} & \st{os6} & \st{os5 v4}\\
\bottomrule
\end{tabular}
\end{adjustbox}
\end{minipage}
\vspace{-2pt}
\caption{Analyzing OS vulnerability for ransomware Locky. Refer to Figure~\ref{fig:cerberOS} for a similar experiment with a different ransomware, same details and conclusions apply to this figure.}
\label{fig:lockyOS}
\vspace{-9pt}
\end{figure}

\textbf{Results.} 
We report the results of these experiments on Cerber and Locky in Figures \ref{fig:cerberOS} and \ref{fig:lockyOS}, respectively. In the tables on the right, those OSs that are correctly discovered by RAPPOR, OLH, and Item-CLDP with correct ranks are depicted in bold. OSs that are correctly discovered but have incorrect rank are depicted in regular font. OSs that privacy methods claim to be among top-10 but in reality are not are depicted with strike-through. We first observe from the tables that the heaviest hitters are correctly discovered in the correct order by all privacy solutions, e.g., top-3 in Cerber. However, as we move lower in the ranking, LDP/CLDP methods start making errors. Particularly for Cerber, RAPPOR and OLH correctly identify only 4 and 5 out of 10 most frequent OSs, respectively, whereas Item-CLDP can identify 9 out of 10. Note that Item-CLDP is missing only the lowest ranked OS (10th), which is arguably the least significant among all ten. 

L1 errors in the graphs on the left show that when $\varepsilon$ is small (0.5 or 1), LDP is competitive against CLDP in this case study. When $\varepsilon$ is higher, CLDP clearly dominates in terms of accuracy. Comparing tabular rankings with L1 scores, we see that CLDP can preserve relative rankings even when its L1 errors are similar to those of LDP. For example, the L1 errors of RAPPOR, OLH, and Item-CLDP are similar when $\varepsilon$~=~1. 
However, studying the top-10 tables shows that Item-CLDP is better at identifying frequent OSs than RAPPOR and OLH.

\vspace{2pt}
\noindent
\underline{\textbf{Case Study \#3: Inspecting Suspicious Activity}}
\vspace{1pt}

\textbf{Setup.} 
In this case study, we consider the sequences of security-related event flags raised by Symantec's behavioral detection engine on each client machine. There are 143 different flags signalling various forms of suspicious activity, ranging from process injection to load point modification. When a flag is raised, it is logged on the client machine with a timestamp, as such the collection of the flagged events constitute a \textit{sequence} over a time period. We investigate the accuracy of collecting these event sequences using Sequence-CLDP. We focus on the same 31-day periods we considered in Case Study \#1, and collect locally private event sequences from the same set of machines infected by ransomware Cerber/Locky. Longitudinal analysis of these event sequences enables Symantec to inspect suspicious activities possibly related to the ransomware infection, e.g., chain of anomalous events leading to the infection. This helps in inferring the precursors or consequences of the infection, and Symnatec can update its detection engine based on the findings. In total, we have 23,558 and 5,717 sequences for Cerber and Locky, respectively, with lengths between 2 to 30.

We use n-gram analysis by mining the top-k popular bigram and trigram patterns from the sequences~\cite{mehdi2009imad,karampatziakis2012using}. We mine the actual patterns that would be obtained if no privacy were applied, and the patterns obtained after Sequence-CLDP is applied. Let $A$ denote the set of actual top-k patterns and $B$ denote the set of top-k patterns mined from perturbed data. We measure their similarity using the Jaccard index: $\text{Jaccard}(A,B) = \frac{|A \cap B|}{|A \cup B|}$. Jaccard similarity is between 0 and 1, with values close to 1 indicating higher similarity. We do not compare against RAPPOR, OLH or SVIM in this case study, since they are not compatible with sequence perturbation.

\begin{figure}
  \centering
  \includegraphics[width=0.88\linewidth]{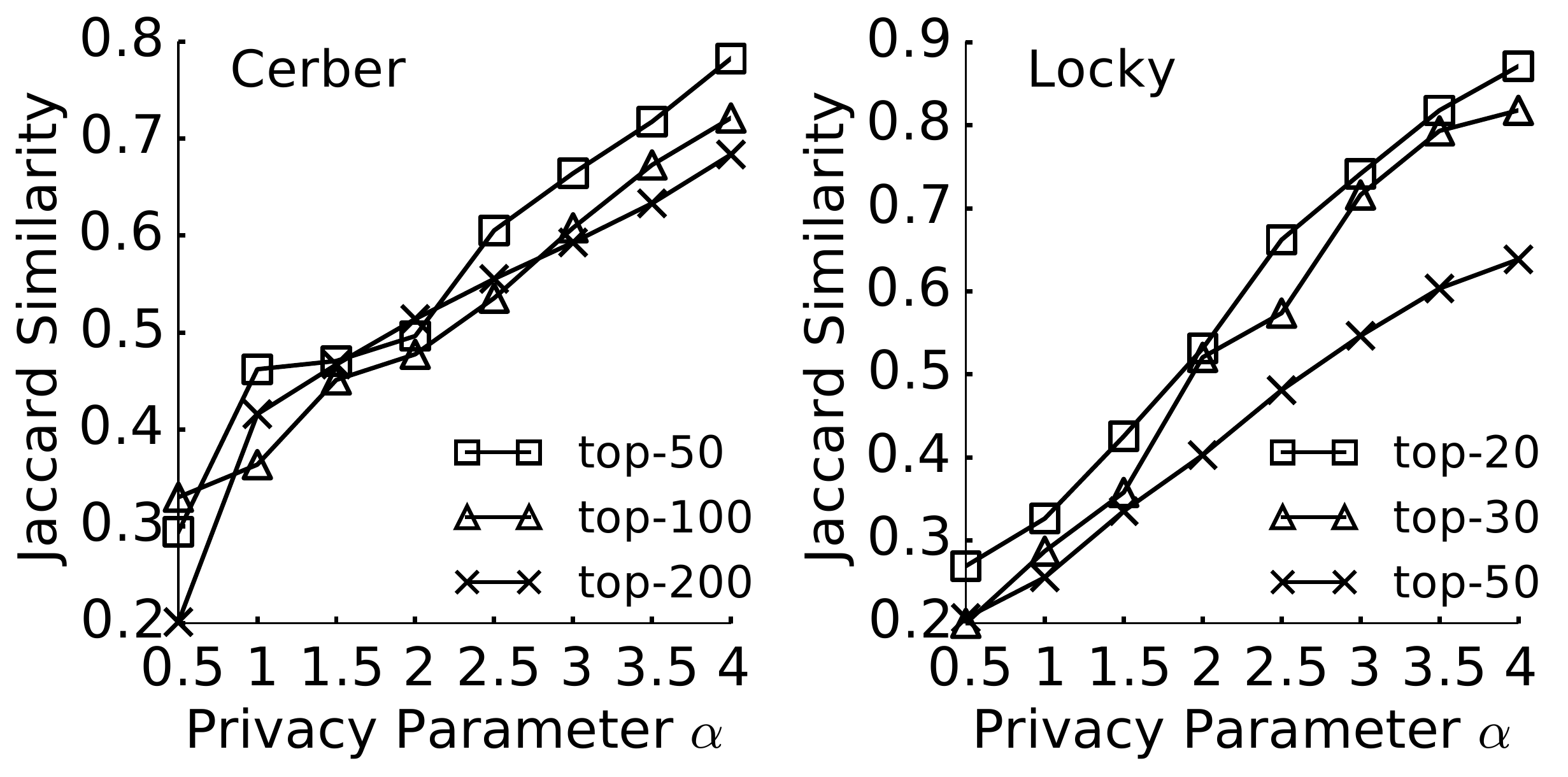} 
  \begin{minipage}[!b]{0.6\linewidth}
  \end{minipage}%
  \begin{minipage}[!b]{0.4\linewidth}
\end{minipage}
\vspace{-6pt}
\caption{
Utility preservation of Sequence-CLDP in mining top-k n-gram patterns from security event sequences obtained from ransomware infected machines. Sequence-CLDP allows discovery of a high fraction of frequent patterns, which are useful to analyze suspicious activity on the machines.}
\label{fig:sequence}
\vspace{-7pt}
\end{figure}

\textbf{Results.} 
The results are shown in Figure \ref{fig:sequence}. We make two important observations. First, as we relax the privacy requirement by increasing $\alpha$, n-grams mined from perturbed sequences become more accurate, as implied by the increase in Jaccard similarity. Second, mining fewer top-k patterns is easier than mining many patterns in general. For example, top-20 in Locky has higher Jaccard similarity score than top-30 and top-50. Similar observation applies to Cerber. This shows Sequence-CLDP preserves the heaviest hitters best, and has higher probability of making errors as n-grams become less and less frequent, which agree with our intuition from Case Study \#2. Note that we mine more patterns in the case of Cerber (up to top-200 as opposed to top-50 for Locky) since the Cerber dataset has more input sequences, thus we can find more n-grams with significant support and confidence.

\vspace{-4pt}
\subsection{Experiments with Public Datasets} \label{sec:public_datasets}

\begin{figure*}[!ht]
\centering
\begin{minipage}[b]{0.46\textwidth}
\includegraphics[width=\textwidth]{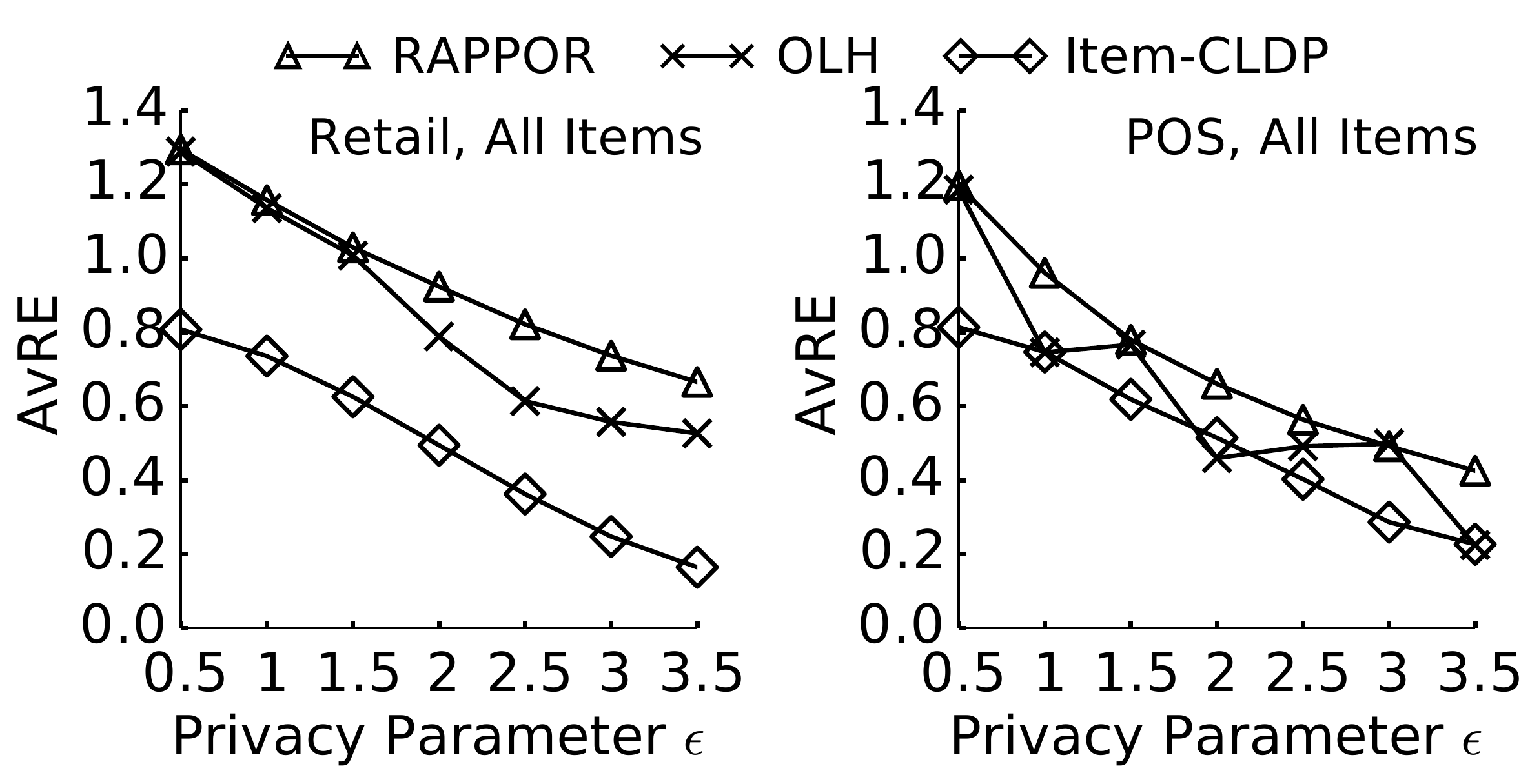} 
\end{minipage}%
\hspace{3mm}
\begin{minipage}[b]{0.46\textwidth}
\includegraphics[width=\textwidth]{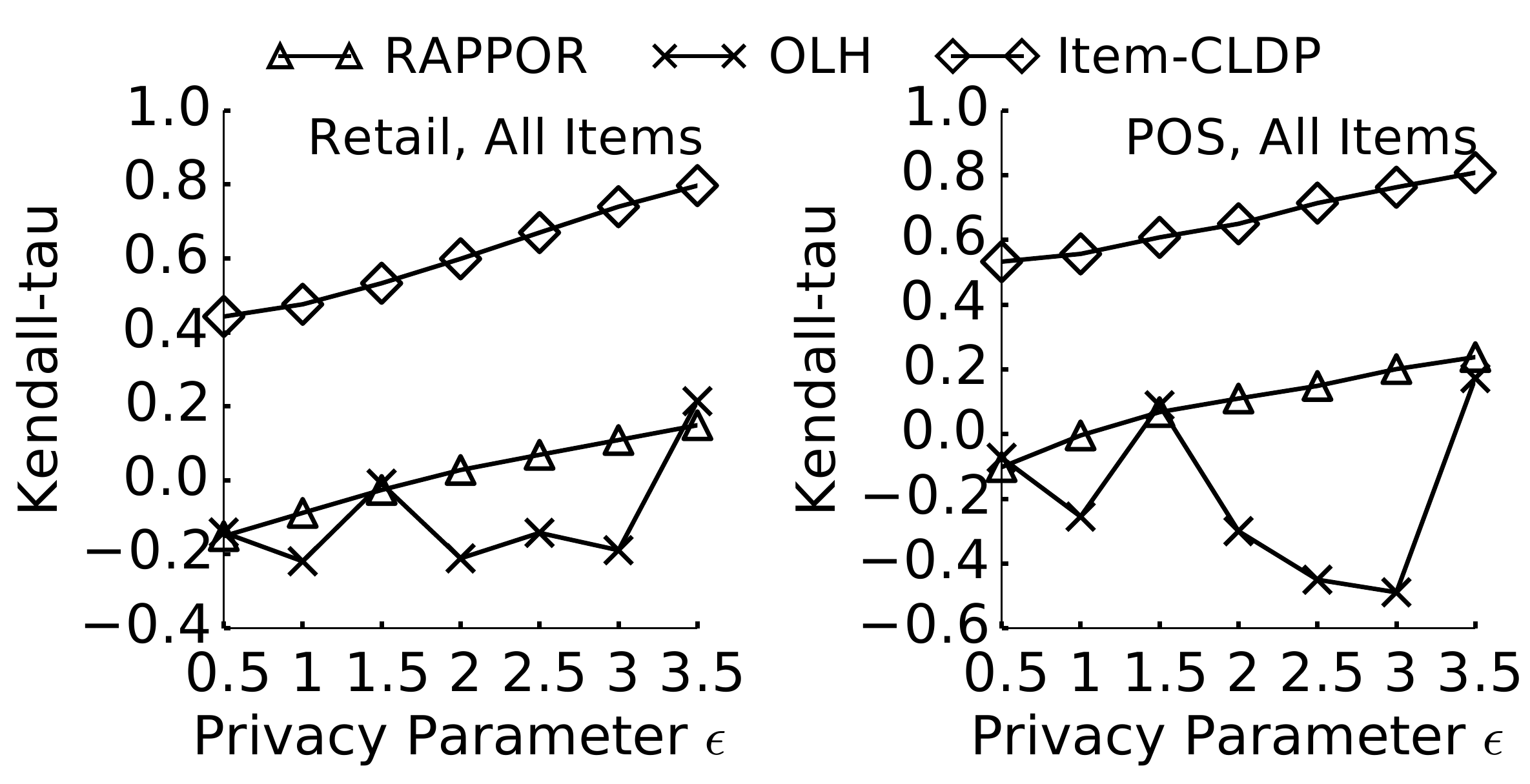}
\end{minipage}%
\vspace{-6pt}
\caption{
AvRE (lower is better) and Kendall-tau scores (higher is better) for singleton experiments across all items in datasets Retail and POS. Item-CLDP provides high utility in estimating item frequencies and rankings across a spectrum of privacy parameter settings.}
\label{fig:singleton_allitems}
\vspace{-11pt}
\end{figure*}

\begin{figure*}[!ht]
\centering
\begin{minipage}[b]{0.46\textwidth} 
\includegraphics[width=\textwidth]{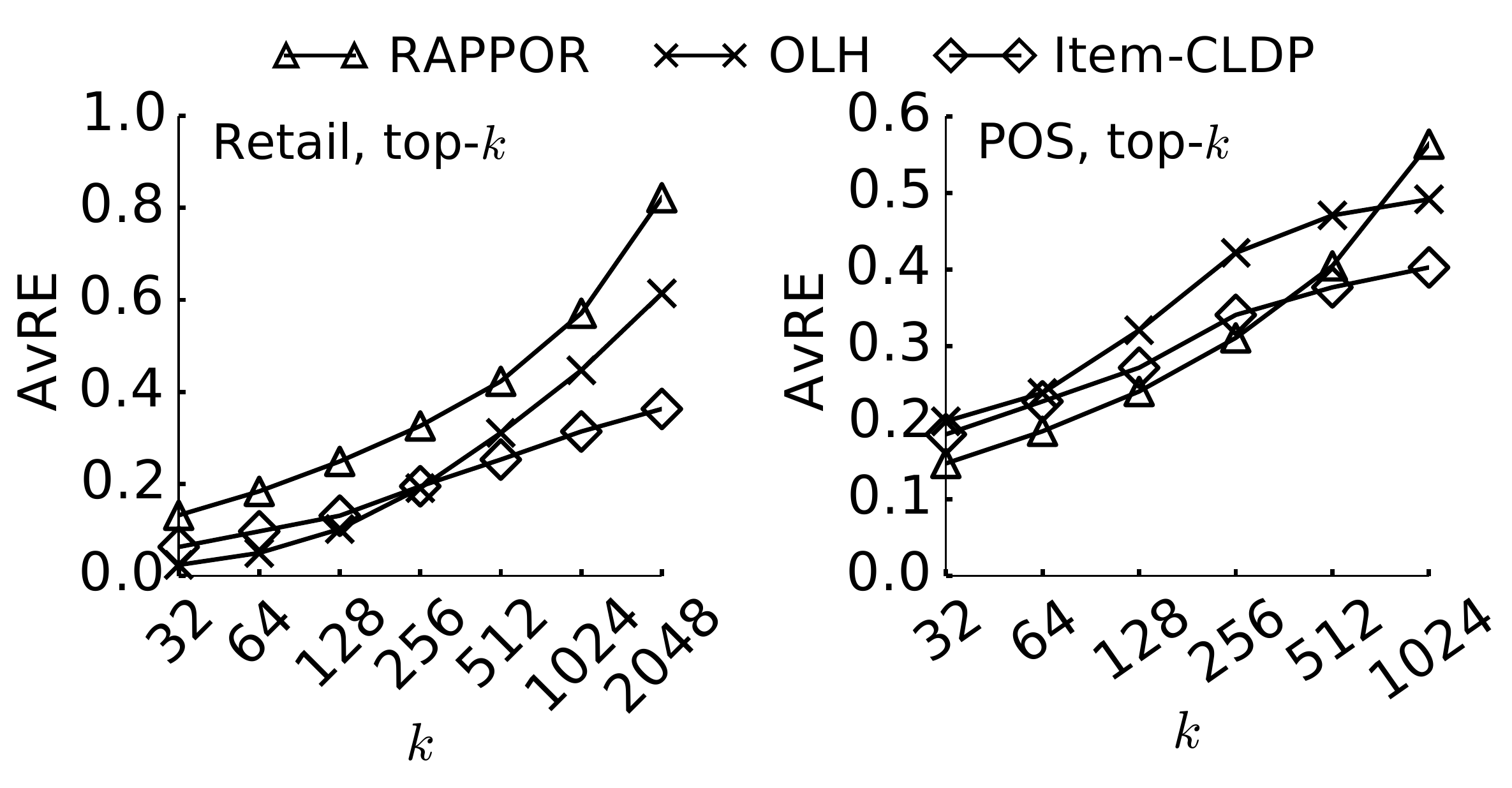} 
\end{minipage}%
\hspace{3mm}
\begin{minipage}[b]{0.46\textwidth} 
\includegraphics[width=\textwidth]{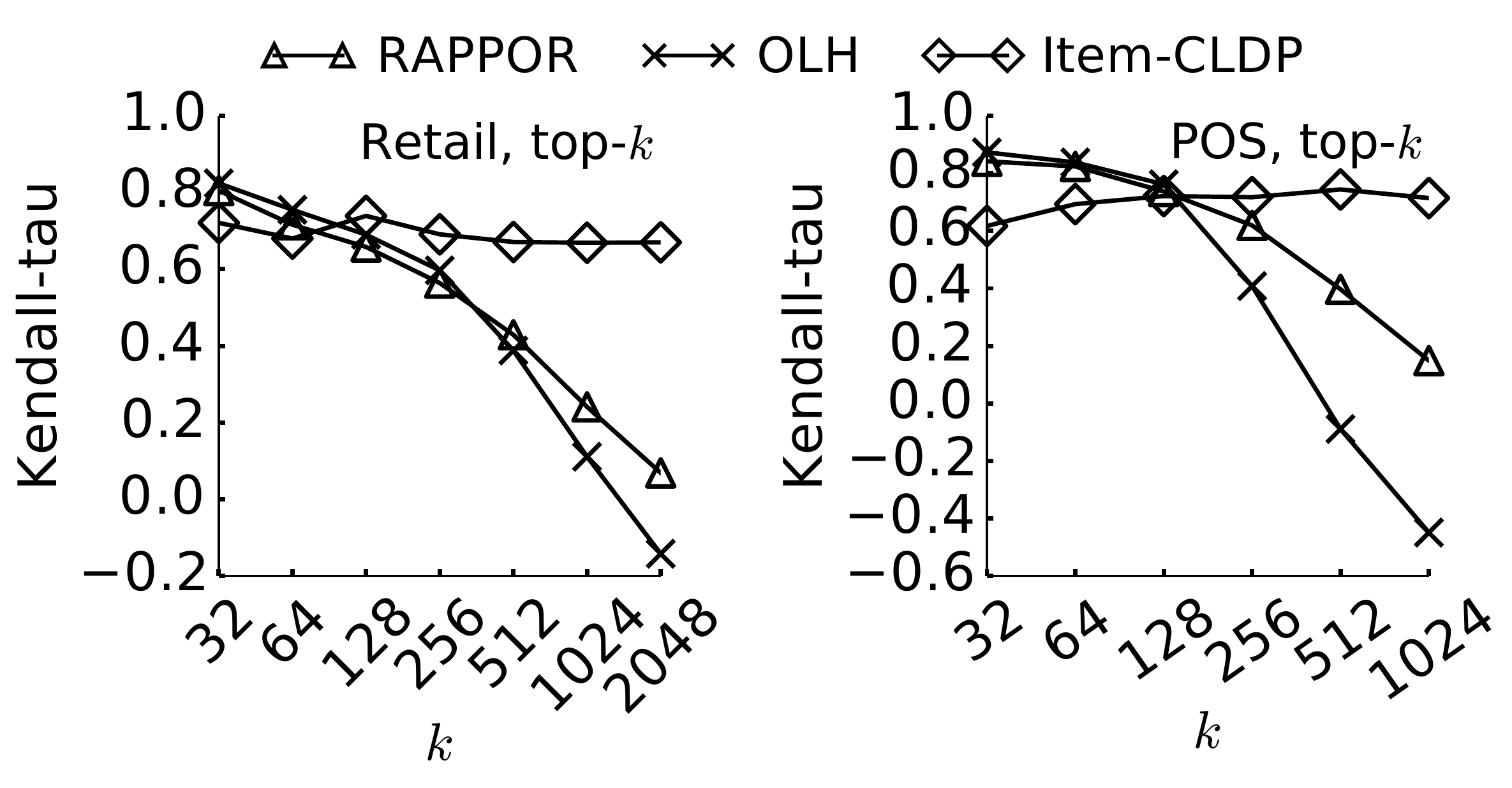}
\end{minipage}%
\vspace{-9pt}
\caption{AvRE and Kendall-tau scores for top-k singletons in Retail and POS. $\varepsilon$ fixed to 2.5, varying $k$ on x-axis. Item-CLDP's accuracy is higher than LDP protocols especially for medium-frequency and infrequent items, e.g., $k \geq$~256.}
\label{fig:singleton_topk}
\vspace{-8pt}
\end{figure*}

\noindent
\textbf{Datasets.} We also experimented on two public datasets: POS and Retail. Both are set-valued datasets. We use them to run singleton non-ordinal item experiments as well as set-valued experiments. For the former, we randomly sample an item from each itemset to create a singleton item dataset. For the latter, we run the set-valued adaptation of our Sequence-CLDP protocol and compare it against SVIM, the state-of-the-art set-valued LDP protocol \cite{wang2018locally}.

\textit{POS} contains several years of market basket sale data from a large electronics retailer \cite{pos-dataset}. It consists of a total of 515,596 transactions with 1,657 unique items sold.

\textit{Retail} contains transactions occurring between January 2010 and September 2011 for a UK-based online retail site \cite{chen2012data}. After cleaning empty entries, this dataset consists of a total of 540,455 transactions with 2,603 unique items.

\vspace{2pt}
\noindent
\textbf{Evaluation Metrics.} We use the following metrics to evaluate the accuracy of the privacy protocols. Similar to our notation from Section \ref{sec:itemCLDP}, let $x$ denote an item, $\text{true}(x)$ denote its true frequency, and $\text{est}(x)$ denote its frequency estimated by the privacy protocol. Let $\mathbf{X}_{gt} = \{x_1, x_2, ..., x_k\}$ be the ground truth top-k items where $x_j$ is the j'th most frequent item. 

\textit{Average Relative Error} (AvRE) measures the mean relative error in top-k items' estimated frequencies versus their true frequencies. Formally:
\[
\text{AvRE} = \frac{\sum_{x \in \mathbf{X}_{gt}} \frac{abs(\text{est}(x)-\text{true}(x))}{\text{true}(x)} }{k} 
\]

The \textit{Kendall-tau coefficient} (KT) measures how well the rankings of heavy hitter top-k items are preserved. A pair of items $x$, $y$ $\in \mathbf{X}_{gt}$ are said to be concordant if their sorted popularity ranks agree, i.e., either of the following hold:
\begin{align*}
    \text{true}(x) > \text{true}(y) &\wedge \text{est}(x) > \text{est}(y) \\
    \text{true}(x) < \text{true}(y) &\wedge \text{est}(x) < \text{est}(y)
\end{align*}
They are said to be discordant if neither holds. Then, the Kendall-tau coefficient of correlation can be defined as:
\[
\text{KT} = \frac{\text{(\# of concordant pairs)}-\text{(\# of discordant pairs)}}{k(k-1)/2}
\]

\vspace{2pt}
\noindent
\textbf{Results of Singleton Experiments.} We run experiments in the singleton setting and compare Item-CLDP with LDP protocols RAPPOR and OLH. Each experiment is repeated 20 times and results are averaged. In Figure \ref{fig:singleton_allitems}, we measure AvRE and Kendall-tau across \textit{all} items by setting $k = |\mathcal{U}|$. Results show that as privacy is relaxed (i.e., $\varepsilon$ and $\alpha$ increase) AvRE decreases and Kendall-tau increases. In most cases, Item-CLDP provides better accuracy; most noticeably, Kendall-tau scores of Item-CLDP are much higher than those of RAPPOR and OLH. When $\varepsilon \geq$~3.5, Kendall-tau scores indicate almost perfect correlation between actual item rankings and rankings found by Item-CLDP, confirming its high accuracy.

Next, we fix the privacy parameter to $\varepsilon=$~2.5 and vary the $k$ (for top-k) to analyze how the protocols behave with respect to varying popularities of items. The results of this experiment are reported in Figure \ref{fig:singleton_topk}. For small $k$ such as $k \leq$~64, there is usually one LDP protocol at least comparable to or better than Item-CLDP. Note that $k=$~64 is a constrained setting covering less than only 6\% of the items in the universe. LDP protocols are optimized to discover such heavy hitters and therefore, they deliver good results when $k$ is small. However, for larger $k$, we observe that Item-CLDP can significantly outperform RAPPOR and OLH. In particular, for $k \geq$~512, under LDP protocols there is almost no correlation in frequency rankings (implied by Kendall-tau results near or below 0), whereas in Item-CLDP, a strong correlation is maintained across all $k$. In short, if the goal is to discover only the top few heavy hitters, LDP protocols offer sufficient accuracy. However, if the goal is to find statistics regarding medium-frequency or infrequent items as well, LDP protocols have inadequate accuracy and we recommend using Item-CLDP.

\begin{figure}[!t]
    \centering
    \includegraphics[width=.47\textwidth]{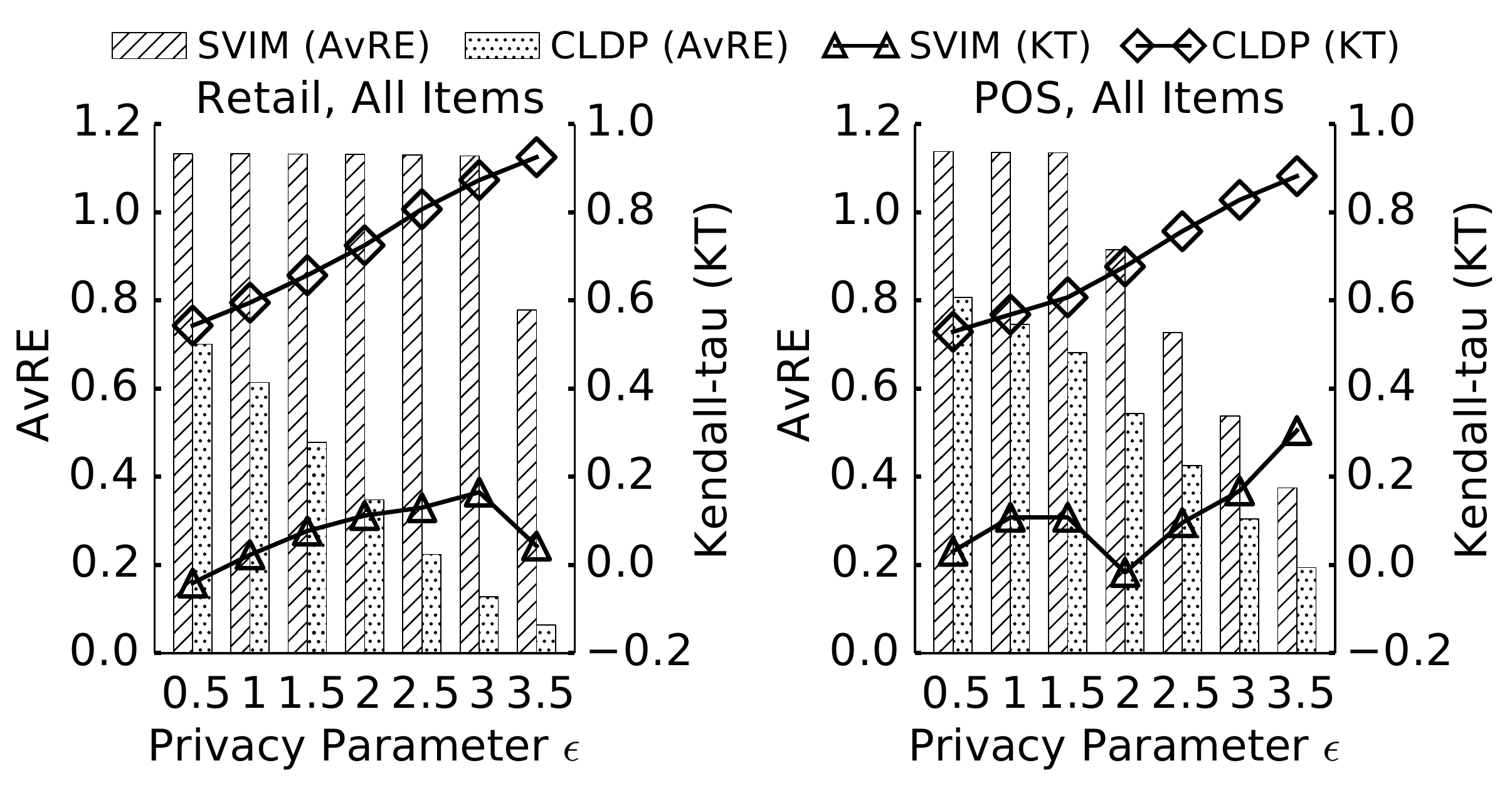}
    \vspace{-4pt}
    \caption{AvRE (bars, left y-axis) and Kendall-tau scores (lines, right y-axis) for set-valued experiments across all items. Sequence-CLDP preserves item frequencies and rankings more effectively than SVIM.} 
    \label{fig:setvalued_allitems}
\end{figure}

\begin{figure}[!t]
    \centering
    \includegraphics[width=.47\textwidth]{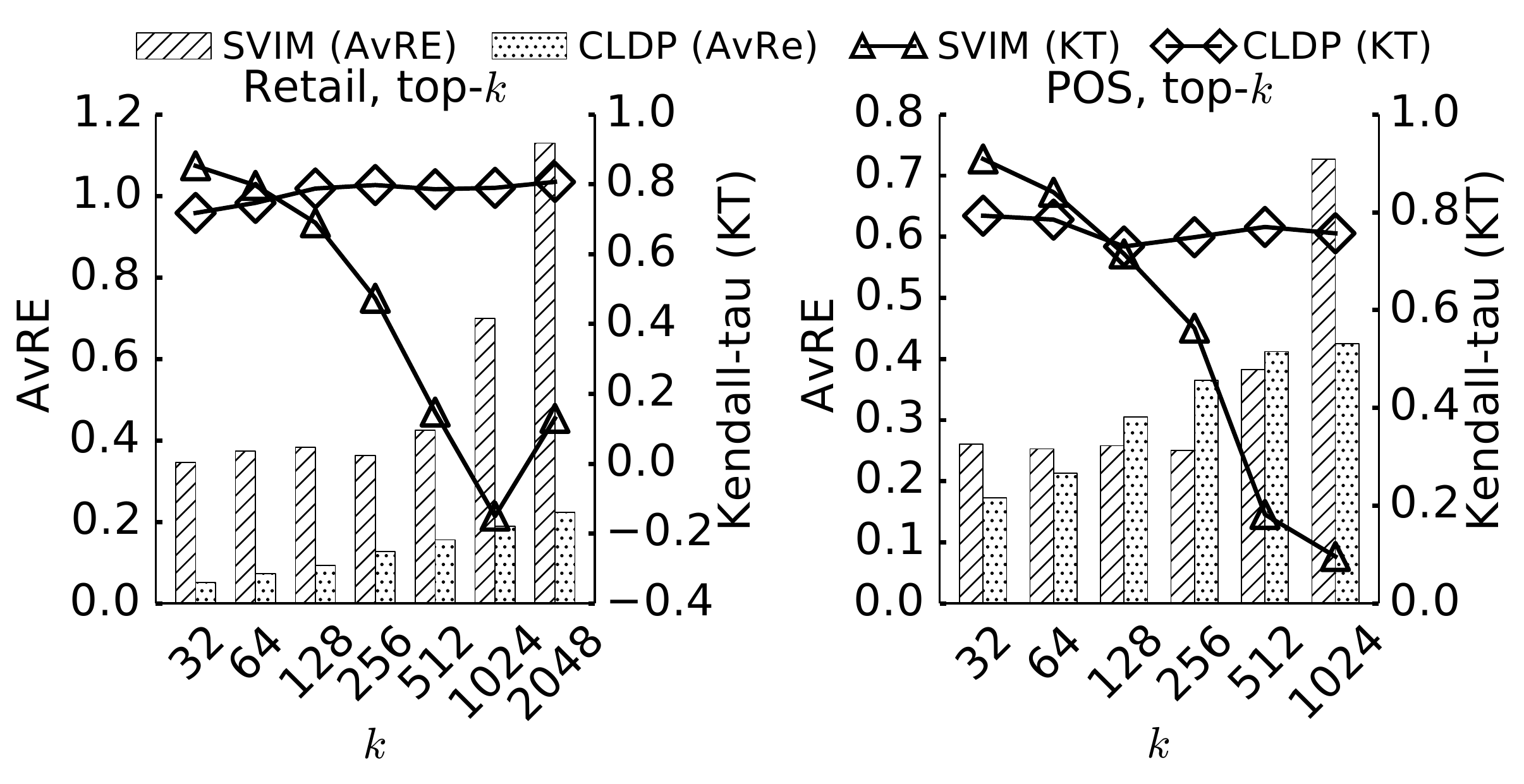}
    \vspace{-4pt}
    \caption{AvRE (bars, left y-axis) and Kendall-tau scores (lines, right y-axis) for set-valued experiments over top-k items with $\varepsilon=$~2.5. Sequence-CLDP offers higher accuracy in terms of AvRE and Kendall-tau for majority of $k$ values.} 
    \label{fig:setvalued_topk}
\end{figure}

\vspace{2pt}
\noindent
\textbf{Results of Set-Valued Experiments.} 
We compare the set-valued adaptation of Sequence-CLDP against the SVIM protocol satisfying LDP \cite{wang2018locally}. Similar to the singleton experiment, we start by setting $k = |\mathcal{U}|$ and vary $\varepsilon$ to study the impact of the privacy budget on accuracy across all items. From the results in Figure \ref{fig:setvalued_allitems}, we observe that Sequence-CLDP offers significant accuracy improvement in terms of both AvRE and Kendall-tau score. For example, the accuracy improvement in terms of AvRE ranges between 30-90\% depending on the dataset and the value of the privacy parameter. In Figure \ref{fig:setvalued_topk}, we fix $\varepsilon$ to 2.5 and vary $k$. As we increase $k$, the accuracy of the protocols generally decrease, since making estimations regarding infrequent items is often more difficult than estimating only the heavy hitters. For Sequence-CLDP, this accuracy decrease is linear or sub-linear; but for SVIM, when $k >$~512, errors start increasing almost exponentially, reiterating that LDP protocols can be poorly suited to estimate statistics regarding infrequent items. Studying the Kendall-tau scores, for very small $k$, Kendall-tau of SVIM and Sequence-CLDP are similar, whereas when $k >$~128, Sequence-CLDP's Kendall-tau scores are significantly better.

\vspace{-7pt}
\section{Related Work}

Differential privacy was initially proposed in the centralized setting in which a trusted central data collector possesses a database containing clients' true values, and noise is applied on the database or queries executed on the database instead of each client's individual value \cite{dwork2006differential, inan2017sensitivity}. In contrast, in LDP, each client locally perturbs their data on their device before sending the perturbed version to the data collector \cite{duchi2013local}. The local setting has seen practical real-world deployment, including Google's RAPPOR as a Chrome extension \cite{erlingsson2014rappor,fanti2016building}, Apple's use of LDP for spelling prediction and emoji frequency detection \cite{thakurta2017emoji,thakurta2017learning}, and Microsoft's collection of application telemetry \cite{ding2017collecting}.

LDP has also sparked interest from the academic community. There have been several theoretical treatments for finding upper and lower bounds on the accuracy and utility of LDP \cite{duchi2013local, kairouz2014extremal, bassily2015local, smith2017interaction, bun2018heavy}. From a more practical perspective, Wang et al.~\cite{wang2017locally} showed the optimality of OLH for singleton item frequency estimation. Qin et al.~\cite{qin2016heavy} and Wang et al.~\cite{wang2018locally} studied frequent item and itemset mining from set-valued client data. Cormode et al.~\cite{cormode2018marginal} and Zhang et al.~\cite{zhang2018calm} studied the problem of obtaining marginal tables from high-dimensional data. Recently, LDP was considered in the contexts of geolocations \cite{chen2016private}, decentralized social graphs \cite{qin2017generating}, and discovering emerging terms from text \cite{wang2018privtrie}.

However, there have also been criticisms and concerns regarding the utility of LDP, which motivated recent works proposing relaxations or alternatives to LDP. BLENDER \cite{avent2017blender} proposed a hybrid privacy model in which only a subset of users enjoy LDP, whereas remaining users act as opt-in beta testers who receive the guarantees of centralized DP. In contrast, our work stays purely in the local privacy model without requiring a trusted data collector (necessary in centralized or hybrid DP) or opt-in clients. Personalized LDP, a weaker form of LDP, was proposed for spatial data aggregation in \cite{chen2016private}; whereas the Restricted LDP scheme proposed in \cite{murakami2018restricted} treats certain client data as more sensitive than others and suggests restricted perturbation schemes to specifically address the more sensitive data. In contrast, our CLDP approach treats all users' data as sensitive (parallel with LDP assumptions) and remains agnostic and extensible with respect to data types.

\vspace{-7pt}
\section{Conclusion}

In this paper, we proposed \textit{Condensed Local Differential Privacy (CLDP)} for utility-aware and privacy-preserving data collection, and developed three protocols: Ordinal-CLDP, Item-CLDP, and Sequence-CLDP for handling a variety of data types prevalent in the cybersecurity domain. Our protocols remain accurate for populations that are an order of magnitude smaller than those required by existing LDP protocols to give adequate accuracy. Case studies using Symantec datasets and experiments on public datasets demonstrate the utility of our proposed approach.

\vspace{-5pt}
\section*{Acknowledgment}
This research is partially supported by the National Science Foundation under NSF grant SaTC 1564097. The first author acknowledges the internship at Symantec in Summer 2018 and the discussions during that intern with Dr.~Kevin Roundy, Dr.~Chris Gates, Dr.~Daniel Marino, and Dr.~Petros Efstathopoulos. Any opinions, findings, and conclusions or recommendations expressed in this material are those of the author(s) and do not necessarily reflect the views of the National Science Foundation or Symantec.

\ifCLASSOPTIONcaptionsoff
  \newpage
\fi

\vspace{-5pt}

\bibliographystyle{IEEEtran}
\vspace{-4pt}

\appendices

\section{Proof of Theorem 1} \label{EMproof}

Recall that for item universe $\mathcal{U}$, Exponential Mechanism (EM), denoted by $\Phi_{EM}$, takes as input true value $v$ and produces fake value $y \in \mathcal{U}$ with probability:
\[
\text{Pr}[\Phi_{EM}(v) = y] = \frac{e^{\frac{-\alpha \cdot d(v,y)}{2}}}{\sum_{z \in \mathcal{U}} e^{\frac{-\alpha \cdot d(v,z)}{2}}}
\]
We prove here that $\Phi_{EM}$ satisfies $\alpha$-CLDP by showing that:
\[
\frac{\text{Pr}[\Phi_{EM}(v_1) = y]}{\text{Pr}[\Phi_{EM}(v_2) = y]} \leq e^{\alpha \cdot d(v_1,v_2)}
\]

\begin{proof}
We start by applying the definition of EM and breaking the odds ratio into two terms:
\begin{align}
\frac{\text{Pr}[\Phi_{EM}(v_1) = y]}{\text{Pr}[\Phi_{EM}(v_2) = y]} &= \frac{\frac{e^{\frac{-\alpha \cdot d(v_1,y)}{2}}}{\sum_{z \in \mathcal{U}} e^{\frac{-\alpha \cdot d(v_1,z)}{2}}}}{\frac{e^{\frac{-\alpha \cdot d(v_2,y)}{2}}}{\sum_{z \in \mathcal{U}} e^{\frac{-\alpha \cdot d(v_2,z)}{2}}}}\\[1ex] \label{eq:stardoublestar}
&= \underbrace{\frac{e^{\frac{-\alpha \cdot d(v_1,y)}{2}}}{e^{\frac{-\alpha \cdot d(v_2,y)}{2}}}}_\text{*} \cdot \underbrace{\frac{\sum_{z \in \mathcal{U}} e^{\frac{-\alpha \cdot d(v_2,z)}{2}}}{\sum_{z \in \mathcal{U}} e^{\frac{-\alpha \cdot d(v_1,z)}{2}}}}_\text{**}
\end{align}
For *, we observe that:
\begin{align}
\frac{e^{\frac{-\alpha \cdot d(v_1,y)}{2}}}{e^{\frac{-\alpha \cdot d(v_2,y)}{2}}} = e^{\frac{\alpha \cdot \big(d(v_2,y)-d(v_1,y)\big)}{2}}
\end{align}
Since $d$ is a metric, it satisfies the triangle inequality. Therefore, it holds that: $d(v_2,y)-d(v_1,y) \leq d(v_1,v_2)$. Combining this with the above, we conclude for *:
\begin{align}
\frac{e^{\frac{-\alpha \cdot d(v_1,y)}{2}}}{e^{\frac{-\alpha \cdot d(v_2,y)}{2}}} \leq e^{\frac{\alpha \cdot d(v_1,v_2)}{2}}
\end{align}
Next, we study the second term **:
\begin{align}
\frac{\sum_{z \in \mathcal{U}} e^{\frac{-\alpha \cdot d(v_2,z)}{2}}}{\sum_{z \in \mathcal{U}} e^{\frac{-\alpha \cdot d(v_1,z)}{2}}} &= \frac{\sum_{z \in \mathcal{U}} e^{\frac{-\alpha \cdot d(v_2,z) + \alpha \cdot d(v_1,z) - \alpha \cdot d(v_1,z)}{2}}}{\sum_{z \in \mathcal{U}} e^{\frac{-\alpha \cdot d(v_1,z)}{2}}}
\end{align}
Again by triangle inequality: $d(v_1,z)-d(v_2,z) \leq d(v_1,v_2)$. Applying this to the numerator we get:
\begin{align}
\frac{\sum_{z \in \mathcal{U}} e^{\frac{-\alpha \cdot d(v_2,z)}{2}}}{\sum_{z \in \mathcal{U}} e^{\frac{-\alpha \cdot d(v_1,z)}{2}}} &\leq \frac{\sum_{z \in \mathcal{U}} e^{\frac{\alpha \cdot d(v_1,v_2) - \alpha \cdot d(v_1,z)}{2}}}{\sum_{z \in \mathcal{U}} e^{\frac{-\alpha \cdot d(v_1,z)}{2}}}\\[1ex]
&\leq \frac{e^{\frac{\alpha \cdot d(v_1,v_2)}{2}} \cdot \sum_{z \in \mathcal{U}} e^{\frac{-\alpha \cdot d(v_1,z)}{2}}}{\sum_{z \in \mathcal{U}} e^{\frac{-\alpha \cdot d(v_1,z)}{2}}}\\[1ex]
&\leq e^{\frac{\alpha \cdot d(v_1,v_2)}{2}}
\end{align}
We established that $* \leq e^{\frac{\alpha \cdot d(v_1,v_2)}{2}}$ and $** \leq e^{\frac{\alpha \cdot d(v_1,v_2)}{2}}$. Plugging them into Equation \ref{eq:stardoublestar} concludes our proof:
\begin{align*}
\frac{\text{Pr}[\Phi_{EM}(v_1) = y]}{\text{Pr}[\Phi_{EM}(v_2) = y]} \leq e^{\frac{\alpha \cdot d(v_1,v_2)}{2}} \cdot e^{\frac{\alpha \cdot d(v_1,v_2)}{2}} = e^{\alpha \cdot d(v_1,v_2)}
\end{align*}
\end{proof}

\section{Utility Properties of Ordinal-CLDP} \label{ordinalproofs}

In Section 4.1 of the main text, we claimed that when $d$ is selected with the property that $d(v_i,v_j)=c$ for $v_i \neq v_j$, where $c$ is a constant, then the aggregation of the outputs of Algorithm 1 in the main text (denoted $\Phi_{ORD}$) preserves the relative frequency rank of $v_i$ and $v_j$ in expectation. Note that since $d$ is a metric, by definition, $d(v_i,v_i)=0$ for all $v_i \in \mathcal{U}$. 

In order to prove this claim, we first introduce some notation. Let $v_i \in \mathcal{U}$ be an item, let $\text{true}(v_i)$ be the true count of $v_i$ in the population, and let $\text{obs}(v_i)$ be the observed count of $v_i$ from the aggregation of perturbed outputs of Algorithm 1. Then, the above claim is equivalent to the statement that iff $\text{true}(v_i) > \text{true}(v_j)$, then $\EX[\text{obs}(v_i)] > \EX[\text{obs}(v_j)]$. Given this holds for all $v_i, v_j \in \mathcal{U}$, the converse follows trivially that iff $\text{true}(v_i) < \text{true}(v_j)$, then $\EX[\text{obs}(v_i)] < \EX[\text{obs}(v_j)]$. Also, if the relative frequency rank relation between all pairs of individual items are preserved in expectation, it implies that that the complete ranking is also preserved. Thus, it suffices to prove the initial statement.

We begin the proof by expressing $\EX[\text{obs}(v_i)]$:
{\small
\begin{align*}
\EX[\text{obs}(v_i)] =~ &\text{true}(v_i) \cdot \text{Pr}[\Phi_{ORD}(v_i) = v_i] \\&+ \text{true}(v_j) \cdot \text{Pr}[\Phi_{ORD}(v_j) = v_i] \\&+ \sum_{v_k \in \mathcal{U} \setminus \{v_i,v_j\}} \text{true}(v_k) \cdot \text{Pr}[\Phi_{ORD}(v_k) = v_i]
\end{align*}
}
Similarly, $\EX[\text{obs}(v_j)]$ can be expressed as:
{\small
\begin{align*}
\EX[\text{obs}(v_j)] =~ &\text{true}(v_j) \cdot \text{Pr}[\Phi_{ORD}(v_j) = v_j] \\&+ \text{true}(v_i) \cdot \text{Pr}[\Phi_{ORD}(v_i) = v_j] \\&+ \sum_{v_k \in \mathcal{U} \setminus \{v_i,v_j\}} \text{true}(v_k) \cdot \text{Pr}[\Phi_{ORD}(v_k) = v_j]
\end{align*}
}

For the above definition of $d$, we observe by construction of $\Phi_{ORD}$ in the main text that $\text{Pr}[\Phi_{ORD}(v_i) = v_i] = \text{Pr}[\Phi_{ORD}(v_j) = v_j]$, $\text{Pr}[\Phi_{ORD}(v_j) = v_i] = \text{Pr}[\Phi_{ORD}(v_i) = v_j]$, and $\text{Pr}[\Phi_{ORD}(v_k) = v_i] = \text{Pr}[\Phi_{ORD}(v_k) = v_j]$. Let constants $c_1$, $c_2$, and $c_3$ denote these three quantities respectively. Then, $\EX[\text{obs}(v_i)] - \EX[\text{obs}(v_j)]$ can be written as:
{\small
\begin{align*}
\EX&[\text{obs}(v_i)] - \EX[\text{obs}(v_j)] =~ c_1 \cdot ( \text{true}(v_i) - \text{true}(v_j)) \\ &- c_2 \cdot (\text{true}(v_i) - \text{true}(v_j)) + c_3 \cdot \sum_{\mathclap{v_k \in \mathcal{U} \setminus \{v_i, v_j\}} } (\text{true}(v_k) - \text{true}(v_k))
\end{align*}
}
which can be simplified to:
{\small
\begin{align} \label{eq:obstrue}
\EX[\text{obs}(v_i)] - \EX[\text{obs}(v_j)] =~ (c_1-c_2) \cdot (\text{true}(v_i) - \text{true}(v_j))
\end{align}
}
By construction of $\Phi_{ORD}$, we know that $c_1 > c_2$. Then, Equation \ref{eq:obstrue} shows that $\EX[\text{obs}(v_i)] - \EX[\text{obs}(v_j)]$ and $\text{true}(v_i) - \text{true}(v_j)$ are directly positively correlated. In other words, iff $\text{true}(v_i) > \text{true}(v_j)$, then $\EX[\text{obs}(v_i)] > \EX[\text{obs}(v_j)]$ and vice versa. This concludes the proof.

\section{Composition in Item-CLDP} 

The Item-CLDP protocol from the main text is a two-round protocol, where the first round satisfies $(\alpha \cdot L)$-CLDP with $d$, and the second round satisfies $\alpha \cdot (1-L)$-CLDP with $d'$. We argued that the composition of the two rounds satisfies:
\begin{align*}
\frac{\text{Pr}[\Phi_{ITEM}(v_1) = \langle v', v'' \rangle]}{\text{Pr}[\Phi_{ITEM}(v_2) = \langle v', v'' \rangle ]} &\leq e^{\alpha \cdot L \cdot d(v_1,v_2)} \cdot e^{\alpha \cdot (1-L) \cdot d'(v_1,v_2)} \\
&\leq e^{\alpha \cdot \max\{d(v_1,v_2), d'(v_1,v_2)\}}
\end{align*}
which is enabled by the fact that $d^*(v_1,v_2) = \max\{d(v_1,v_2), d'(v_1,v_2)\}$ is also a metric, given that $d$ and $d'$ are metrics by their construction within Item-CLDP. 

Here, we prove this fact that enables Item-CLDP composition. That is, given $d(v_i,v_j)$ and $d'(v_i, v_j)$ are metrics, $d^*(v_i,v_j) = \max\{d(v_i,v_j), d'(v_i,v_j)\}$ is also a metric.

First recall the properties of being a metric, which are satisfied by $d$ and $d'$. We write these properties for $d$, as $d'$ is analogous:
\begin{enumerate}
    \item $d(v_i,v_j) \geq 0$, $\forall v_i, v_j \in \mathcal{U}$
    \item $d(v_i, v_j) = 0$ if and only if $v_i = v_j$
    \item $d(v_i, v_j) = d(v_j, v_i)$, $\forall v_i, v_j \in \mathcal{U}$
    \item $d(v_i, v_k) \leq d(v_i, v_j) + d(v_j, v_k)$, $\forall v_i, v_j, v_k \in \mathcal{U}$
\end{enumerate}
To prove that $d^*$ is also a metric, we must show that $d^*$ satisfies the four properties above. We prove each property one by one.

\noindent
\underline{Property 1:} We know already that $d(v_i,v_j) \geq 0$ and $d'(v_i,v_j) \geq 0$ for all $v_i, v_j$. $d^*(v_i,v_j)$ takes the maximum of these two non-negative values, which must also be non-negative. Thus, $d^*(v_i,v_j) \geq 0$. 

\noindent
\underline{Property 2:} There are two directions to prove. First, when $v_i = v_j$ then $d(v_i,v_j)=0$ and $d'(v_i,v_j)=0$. As a result, $d^*(v_i,v_j) = \max\{0, 0\} = 0$, which concludes the first direction. 

For the second direction, we prove by contradiction. Assume that $d^*(v_i,v_j) = 0$, but contrary to expectation $v_i \neq v_j$. If $v_i \neq v_j$, by the first and second properties satisfied by $d$ and $d'$, either $d(v_i,v_j) > 0 $ or $d'(v_i, v_j) > 0$ must hold. However, if either holds, the maximum of $d(v_i,v_j)$ and $d'(v_i, v_j)$ cannot be zero, since at least one of them is positive. Thus, the initial assumption of $d^*(v_i,v_j) = 0$ is violated, which yields a contradiction. It must be that $v_i = v_j$. 

\noindent
\underline{Property 3:} The symmetry property of $d^*$ follows from the individual symmetry properties satisfied by $d$ and $d'$.
\begin{align*}
    d^*(v_i,v_j) &= \max\{d(v_i,v_j), d'(v_i,v_j)\} \\
                &= \max\{d(v_j,v_i), d'(v_j,v_i)\} \\
                &= d^*(v_j,v_i)
\end{align*}

\noindent
\underline{Property 4:} By definition of $d^*$, $d^*(v_i, v_k)$ is either equal to $d(v_i, v_k)$ or $d'(v_i, v_k)$. Without loss of generality assume $d^*(v_i, v_k) = d(v_i, v_k)$; the case with $d^*(v_i, v_k) = d'(v_i, v_k)$ is identical. Using the fourth property of $d$, we can derive:
\begin{align*}
    d^*(v_i, v_k) &= d(v_i, v_k) \\
                    &\leq d(v_i, v_j) + d(v_j, v_k) \\
                    &\leq d^*(v_i, v_j) + d^*(v_j, v_k)
\end{align*}
since for each individual term on the right hand side, it is guaranteed by definition of $d^*$ that $d^*(v_i, v_j) \geq d(v_i, v_j)$ and $d^*(v_j, v_k) \geq d(v_j, v_k)$.

\section{Proof of Theorem 2} \label{sequenceCLDPproof}

This section contains the privacy proofs for the Sequence-CLDP perturbation mechanism given in Algorithm 2 of the main text. We have claimed in the main text that Sequence-CLDP satisfies two properties: length indistinguishability and content indistinguishability. We organize this section into two subsections for proving each property.

\subsection{Length Indistinguishability}

We first prove that Sequence-CLDP, denoted here onwards by $\Phi_{SEQ}$, satisfies $\alpha$-length-indis\-tinguishability when certain value ranges are enforced for its parameters $\textit{halt}$, $\textit{gen}$. We need to show that for any pair of true sequences $X$, $Y$:
\[
\frac{\text{Pr}[\Phi_{SEQ}(X) \leadsto \ell]} {\text{Pr}[\Phi_{SEQ}(Y) \leadsto \ell]} \leq e^{\alpha \cdot abs(|X|-|Y|)}
\]

\begin{proof}
Observe from Algorithm 2 in the main text that given a true sequence $X$ of length $|X|$, $\Phi_{SEQ}$ produces an output sequence of length $\ell$ with probability:
\[
\text{Pr}[\Phi_{SEQ}(X) \leadsto \ell] = 
\begin{cases}
\textit{halt} \cdot (1-\textit{halt})^{\ell} \\ \hfill \text{ when } \ell < |X| \\ \\
(1-\textit{halt})^{|X|} \cdot (1-\textit{gen}) \cdot \textit{gen}^{\ell-|X|} \\ \hfill \text{ when } \ell \geq |X| 
\end{cases}
\]
We consider several disjoint cases depending on the length of true sequences $X$, $Y$. The parameters must be chosen so that all cases are simultaneously satisfied. 

\noindent
\underline{Case 0: $|X|=|Y|$}. This case is trivial since $\Phi_{SEQ}$ treats their length equally, resulting in:
\[
\frac{\text{Pr}[\Phi_{SEQ}(X) \leadsto \ell]} {\text{Pr}[\Phi_{SEQ}(Y) \leadsto \ell]}  = 1 = e^{\alpha \cdot 0} \qquad \text{ if } |X|=|Y|
\]
Remaining cases fall under $|X| \neq |Y|$, and are analyzed case-by-case below.

\noindent
\underline{Case 1: $\ell < |X|$ and $\ell < |Y|$} 
\[
\frac{\text{Pr}[\Phi_{SEQ}(X) \leadsto \ell]} {\text{Pr}[\Phi_{SEQ}(Y) \leadsto \ell]} = \frac{\textit{halt} \cdot (1-\textit{halt})^{\ell}}{\textit{halt} \cdot (1-\textit{halt})^{\ell}} = 1 \leq e^{\alpha \cdot abs(|X|-|Y|)}
\]
Since $abs(|X|-|Y|) \geq 0$, we trivially have $e^{\alpha \cdot abs(|X|-|Y|)} \geq 1$, hence this case is satisfied without placing constraints on the values of \textit{halt}, \textit{gen}.

\noindent
\underline{Case 2: $\ell \geq |X|$ and $\ell \geq |Y|$}
\begin{align*}
\frac{\text{Pr}[\Phi_{SEQ}(X) \leadsto \ell]} {\text{Pr}[\Phi_{SEQ}(Y) \leadsto \ell]} &= \frac{(1-\textit{halt})^{|X|} \cdot (1-\textit{gen}) \cdot \textit{gen}^{\ell-|X|}}{(1-\textit{halt})^{|Y|} \cdot (1-\textit{gen}) \cdot \textit{gen}^{\ell-|Y|}} \\ &= (1-\textit{halt})^{|X|-|Y|} \cdot \textit{gen}^{|Y|-|X|}
\end{align*}
Divide this into two subcases:

\noindent
\underline{2a: $|X| < |Y|$}: When this holds, the requirement that
\[
(1-\textit{halt})^{|X|-|Y|} \cdot \textit{gen}^{|Y|-|X|} \leq e^{\alpha \cdot abs(|X|-|Y|)}
\]
can be written as:
\begin{align*}
\frac{\textit{gen}^{|Y|-|X|}}{(1-\textit{halt})^{|Y|-|X|}} &\leq e^{\alpha \cdot (|Y|-|X|)} \\
(\frac{\textit{gen}}{1-\textit{halt}})^{|Y|-|X|} &\leq e^{\alpha \cdot (|Y|-|X|)}
\end{align*}
Thus, we have the constraints for parameters \textit{halt}, \textit{gen} as:
\begin{align} \label{eq:cons_2a}
    \frac{\textit{gen}}{1-\textit{halt}} \leq e^{\alpha}
\end{align}

\noindent
\underline{2b: $|X| > |Y|$}: When this holds, the same requirement can be written as:
\begin{align*}
\frac{(1-\textit{halt})^{|X|-|Y|}}{\textit{gen}^{|X|-|Y|}} &\leq e^{\alpha \cdot (|X|-|Y|)} \\ 
(\frac{1-\textit{halt}}{\textit{gen}})^{|X|-|Y|} &\leq e^{\alpha \cdot (|X|-|Y|)}
\end{align*}
resulting in the parameter constraints:
\begin{align} \label{eq:cons_2b}
    \frac{1-\textit{halt}}{\textit{gen}} \leq e^{\alpha}
\end{align}

\noindent
\underline{Case 3: $\ell < |X|$ and $\ell \geq |Y|$}
\begin{align*}
\frac{\text{Pr}[\Phi_{SEQ}(X) \leadsto \ell]} {\text{Pr}[\Phi_{SEQ}(Y) \leadsto \ell]} &\leq e^{\alpha \cdot abs(|X|-|Y|)} \\
\frac{\textit{halt} \cdot (1-\textit{halt})^{\ell}}{(1-\textit{halt})^{|Y|} \cdot (1-\textit{gen}) \cdot \textit{gen}^{\ell-|Y|}} &\leq e^{\alpha \cdot abs(|X|-|Y|)} \\
\frac{\textit{halt}}{1-\textit{gen}} \cdot (\frac{1-\textit{halt}}{\textit{gen}})^{\ell - |Y|} &\leq e^{\alpha \cdot abs(|X|-|Y|)}
\end{align*}
Since the assumption in this case is $|Y| \leq \ell < |X|$, we can rewrite the RHS as:
\[
\frac{\textit{halt}}{1-\textit{gen}} \cdot (\frac{1-\textit{halt}}{\textit{gen}})^{\ell - |Y|} \leq e^{\alpha \cdot (\ell - |Y|)} \cdot e^{\alpha \cdot (|X|-\ell)}
\]
Given the constraint we established in Equation \ref{eq:cons_2b} holds, the following is a sufficient condition to satisfy the above:
\[
\frac{\textit{halt}}{1-\textit{gen}} \leq e^{\alpha \cdot (|X|-\ell)}
\]
Notice that, by assumption, $|X| > \ell$ and both $|X|$ and $\ell$ are integers representing sequence length. Then, $|X|-\ell \geq 1$, making the following parameter constraint sufficient:
\begin{align} \label{eq:cons3}
\frac{\textit{halt}}{1-\textit{gen}} \leq e^{\alpha}
\end{align}

\noindent
\underline{Case 4: $\ell \geq |X|$ and $\ell < |Y|$}
\begin{align*}
\frac{\text{Pr}[\Phi_{SEQ}(X) \leadsto \ell]} {\text{Pr}[\Phi_{SEQ}(Y) \leadsto \ell]} &\leq e^{\alpha \cdot abs(|X|-|Y|)} \\
\frac{(1-\textit{halt})^{|X|} \cdot (1-\textit{gen}) \cdot \textit{gen}^{\ell - |X|}}{\textit{halt} \cdot (1-\textit{halt})^{\ell}} &\leq e^{\alpha \cdot abs(|X|-|Y|)} \\
\frac{1-\textit{gen}}{\textit{halt}} \cdot (\frac{\textit{gen}}{1-\textit{halt}})^{\ell-|X|} &\leq e^{\alpha \cdot abs(|X|-|Y|)}
\end{align*}
Since the assumption in this case is $|X| \leq \ell < |Y|$, we can rewrite the RHS as:
\[
\frac{1-\textit{gen}}{\textit{halt}} \cdot (\frac{\textit{gen}}{1-\textit{halt}})^{\ell-|X|} \leq e^{\alpha \cdot (|Y|-\ell)} \cdot  e^{\alpha \cdot (\ell - |X|)}
\]
Given the constraint we established in Equation \ref{eq:cons_2a} holds, the following is a sufficient condition to satisfy the above:
\[
\frac{1-\textit{gen}}{\textit{halt}} \leq e^{\alpha \cdot (|Y|-\ell)}
\]
Since $|Y| > \ell$ and both $|Y|$ and $\ell$ are integers, we have $|Y|-\ell \geq 1$, making the following parameter constraint sufficient:
\begin{align} \label{eq:cons4}
\frac{1-\textit{gen}}{\textit{halt}} \leq e^{\alpha} 
\end{align}

\noindent
\underline{Combine all cases:} Finally, we combine the parameter constraints we identified at the end of each case (Equations \ref{eq:cons_2a}, \ref{eq:cons_2b}, \ref{eq:cons3}, and \ref{eq:cons4}) to obtain a system of equations:
\begin{align*}
\frac{\textit{gen}}{1-\textit{halt}} &\leq e^{\alpha} \qquad \frac{1-\textit{halt}}{\textit{gen}} \leq e^{\alpha} \\ \frac{\textit{halt}}{1-\textit{gen}} &\leq e^{\alpha} \qquad \frac{1-\textit{gen}}{\textit{halt}} \leq e^{\alpha}
\end{align*}
Given the privacy parameter $\alpha$, the values of \textit{halt}, \textit{gen} satisfying all four equations simultaneously satisfy $\alpha$-length-indistinguishability. If we set \textit{halt}=\textit{gen} and solve this system of equations, we observe that the following is a solution (which we call the symmetric solution in the main text):
\[
\textit{halt} = \textit{gen} = \frac{1}{e^{\alpha}+1}
\]
Another solution to the same system of equations, which we call the asymmetric solution, is desirable when input sequences $X$, $Y$ are longer and therefore a smaller \textit{halt}ing probability is preferable:
\[
0 < \textit{halt} < \frac{1}{e^{\alpha}+1} ~~\text{and}~~ 1 - e^{\alpha} \cdot \textit{halt} \leq \textit{gen} \leq 1 - \frac{\textit{halt}}{e^{\alpha}}
\]
\end{proof}

\subsection{Content Indistinguishability} 

Next, we prove that for the above choice of parameters, $\Phi_{SEQ}$ satisfies $\alpha$-content-indistin\-guishability. That is, for any pair of true sequences $X$, $Y$ of same length, the following holds:
\[
\frac{\text{Pr}[\Phi_{SEQ}(X) = S]} {\text{Pr}[\Phi_{SEQ}(Y) = S]} \leq e^{\alpha \cdot d_{\text{seq}}(X,Y)}
\]

\begin{proof}
We divide into 3 possible cases depending on how $|S|$ relates to $|X|$ = $|Y|$.

\noindent
\underline{Case 1: $|S|=|X|=|Y|$.} In this case, Algorithm 2 must have behaved as follows. Upon reaching its main loop (lines 5-14), it must have run lines 6-9 for $i=1$ to \textit{max\_len} and, in each iteration, the event with probability $(1-\textit{halt})$ must have occurred such that $X[i]$ was perturbed and the perturbed item was appended to $S$. The perturbation is performed by the function call to Algorithm 1, which we denote by $\Phi_{EM}$. Then, in iteration $i=\textit{max\_len}+1$, the event with probability $(1-\textit{gen})$ must have occurred so that a sequence $S$ with length exactly equal to $|X|=|Y|=n$ was returned. 
The odds-ratio probabilities of the overall run can be computed as:

{\footnotesize
\begin{align*}
\frac{\text{Pr}[\Phi_{SEQ}(X) = S]} {\text{Pr}[\Phi_{SEQ}(Y) = S]} &\stackrel{?}{\leq} e^{\alpha \cdot d_{\text{seq}}(X,Y)} \\
\frac{(1-\textit{gen}) \cdot \prod_{i=1}^{n} (1-\textit{halt}) \cdot \text{Pr}[\Phi_{EM}(X[i]) = S[i]]}{(1-\textit{gen}) \cdot \prod_{i=1}^{n} (1-\textit{halt}) \cdot \text{Pr}[\Phi_{EM}(Y[i]) = S[i]]} &\stackrel{?}{\leq} e^{\alpha \cdot d_{\text{seq}}(X,Y)}\\
\frac{\prod_{i=1}^{n} \text{Pr}[\Phi_{EM}(X[i]) = S[i]]}{\prod_{i=1}^{n} \text{Pr}[\Phi_{EM}(Y[i]) = S[i]]} &\stackrel{?}{\leq} e^{\alpha \cdot d_{\text{seq}}(X,Y)}
\end{align*}
}
By Theorem 1, for any index $i$:
\begin{align} \label{eq:bytheorem1}
\frac{\text{Pr}[\Phi_{EM}(X[i]) = S[i]]}{\text{Pr}[\Phi_{EM}(Y[i]) = S[i]]} \leq e^{\alpha \cdot d(X[i],Y[i])}
\end{align}
Applying Equation \ref{eq:bytheorem1}, we obtain:
\begin{align*}
\prod_{i=1}^{n} e^{\alpha \cdot d(X[i],Y[i])} &\stackrel{?}{\leq} e^{\alpha \cdot d_{\text{seq}}(X,Y)} \\
e^{\alpha \cdot \sum_{i=1}^{n} d(X[i],Y[i])} &\leq e^{\alpha \cdot d_{\text{seq}}(X,Y)}
\end{align*}
which holds by definition of $d_{\text{seq}}$, completing proof for Case 1.

\noindent
\underline{Case 2: $|S|<|X|=|Y|$.} In this case, Algorithm 2 must have run lines 6-9 for $|S|+1$ iterations; in the first $|S|$ iterations, the event with probability $(1-\textit{halt})$ must have occurred to build the perturbed sequence $S$, and in the last iteration the \textit{halt}ing event must have occurred so that the sequence of length $|S|=m$ was returned without adding more elements. The odds-ratio becomes:

\begin{align*}
\frac{\text{Pr}[\Phi_{SEQ}(X) = S]} {\text{Pr}[\Phi_{SEQ}(Y) = S]} &\stackrel{?}{\leq} e^{\alpha \cdot d_{\text{seq}}(X,Y)} \\
\frac{\textit{halt} \cdot \prod_{i=1}^{m} (1-\textit{halt}) \cdot \text{Pr}[\Phi_{EM}(X[i]) = S[i]]}{\textit{halt} \cdot \prod_{i=1}^{m} (1-\textit{halt}) \cdot \text{Pr}[\Phi_{EM}(Y[i]) = S[i]]} &\stackrel{?}{\leq} e^{\alpha \cdot d_{\text{seq}}(X,Y)} \\ 
\frac{\prod_{i=1}^{m} \text{Pr}[\Phi_{EM}(X[i]) = S[i]]} {\prod_{i=1}^{m} \text{Pr}[\Phi_{EM}(Y[i]) = S[i]]} &\stackrel{?}{\leq} e^{\alpha \cdot d_{\text{seq}}(X,Y)}
\end{align*}
Notice that this case is different than Case 1 in the bounds of the product---the product goes from $i=1$ to length of $|S|$, which is shorter than $|X|$. Applying Equation \ref{eq:bytheorem1}:
\begin{align*}
e^{\alpha \cdot \sum_{i=1}^{m} d(X[i],Y[i])} \stackrel{?}{\leq} e^{\alpha \cdot d_{\text{seq}}(X,Y)}
\end{align*}
A distance metric $d$ satisfies the non-negativity property by definition. Therefore, for $m < n=|X|$, we have: 
\begin{align*}
\sum_{i=1}^{m} d(X[i],Y[i]) \leq \sum_{i=1}^{n} d(X[i],Y[i]) = d_{\text{seq}}(X,Y)
\end{align*}
completing the proof for Case 2.

\noindent
\underline{Case 3: $|S|>|X|=|Y|$.} In this case, Algorithm 2 must have run lines 6-9 for $|X|$ iterations, and in each iteration the event with probability $(1-\textit{halt})$ must have occurred. Then, for $|S|-|X|$ iterations, lines 10-13 must have run, and the item generation event with probability \textit{gen} must have occurred in these iterations. Finally, to return $S$, a final iteration must have occurred with the stopping event having $(1-\textit{gen})$ probability. Let $m = |S|$ and $n = |X|$. The odds-ratio is: 
\begin{align*}
\frac{\text{Pr}[\Phi_{SEQ}(X) = S]} {\text{Pr}[\Phi_{SEQ}(Y) = S]} &\stackrel{?}{\leq} e^{\alpha \cdot d_{\text{seq}}(X,Y)} \\
\frac{ \begin{array}{@{}r@{}}
(1-\textit{gen}) \cdot \prod_{i=1}^{m-n} \textit{gen} \cdot Pr[\text{random}=S[i]] \\ {}
\cdot \prod_{i=1}^{n} (1-\textit{halt}) \cdot \text{Pr}[\Phi_{EM}(X[i]) = S[i]]
\end{array}
}
{ \begin{array}{@{}r@{}}
(1-\textit{gen}) \cdot \prod_{i=1}^{m-n} \textit{gen} \cdot Pr[\text{random}=S[i]] \\ {}
\cdot \prod_{i=1}^{n} (1-\textit{halt}) \cdot \text{Pr}[\Phi_{EM}(Y[i]) = S[i]]
\end{array}
} &\stackrel{?}{\leq} e^{\alpha \cdot d_{\text{seq}}(X,Y)}
\end{align*}
where $Pr[\text{random}=S[i]]$ denotes the probability that the random sampling on line 11 of Algorithm 2 returns item $S[i]$. Note that the random sampling does not depend on the properties of inputs $X$, $Y$, therefore we can safely cancel most terms. We end up with:
\begin{align*}
\frac{\prod_{i=1}^{n} \text{Pr}[\Phi_{EM}(X[i]) = S[i]]}{\prod_{i=1}^{n} \text{Pr}[\Phi_{EM}(Y[i]) = S[i]]} &\stackrel{?}{\leq} e^{\alpha \cdot d_{\text{seq}}(X,Y)}    
\end{align*}
Applying Equation \ref{eq:bytheorem1} and continuing in the same fashion as Case 1, it is straightforward to complete the proof. 
\end{proof}

\end{document}